\documentclass[epj]{svjour}
\usepackage{url}
\usepackage{hyperref}
\usepackage{xspace}
\usepackage{graphicx}
\usepackage{amssymb}
\usepackage{multirow}
\usepackage{array}
\usepackage[usenames,dvipsnames]{color}
\usepackage{amsmath}
\usepackage[pagewise]{lineno}

\usepackage{graphics}

\journalname{Eur. Phys. J. C}

\begin{document}
\title{Search for the sterile neutrino mixing with the ICAL detector at 
INO}
\author{S. P. Behera$^{1,2,}$\thanks {e-mail:shiba@barc.gov.in}, Anushree 
Ghosh$^{3,}$, Sandhya Choubey$^{4}$, V. M. Datar$^{5}$, D. K. Mishra$^{1}$, 
\and A. K. Mohanty$^{1,2,}$\thanks{Present Address: Saha Institute of 
Nuclear Physics, 1/AF, Bidhan nagar, Kolkata - 700064, India}}

\institute{Nuclear Physics Division, Bhabha Atomic Research Centre, Mumbai 
- 400085, India \and Homi Bhabha National Institute, Mumbai - 400094, India, 
\and  Universidad Tecnica Federico Santa Maria - Departamento de 
Fisica Casilla 110-V, Avda. Espana 1680, Valparaiso, Chile, \and Harish-Chandra 
Research Institute, Chhatnag Road, Jhunsi, Allahabad - 211 019, India, \and INO Cell, Tata 
Institute of Fundamental Research, Mumbai 400005, India}

\abstract{
The study has been carried out on the prospects of probing the sterile neutrino 
mixing with the magnetized Iron CALorimeter (ICAL) at the India-based Neutrino 
Observatory (INO), using atmospheric neutrinos as a source. The so-called 
3~$+$~1 scenario is considered for active-sterile neutrino mixing and lead to 
projected exclusion curves in the sterile neutrino mass and mixing angle plane. 
The analysis is performed using the neutrino event generator NUANCE, modified 
for ICAL, and folded with the detector resolutions obtained by the INO 
collaboration from a full GEANT4 based detector simulation. A comparison has 
been made between the results obtained from the analysis considering only the 
energy and zenith angle of the muon and combined with the hadron energy due to 
the neutrino induced event. A small improvement has been observed with the 
addition of the hadron information to the muon. In the analysis we consider 
neutrinos coming from all zenith angles and the Earth matter effects are also 
included. The inclusion of events from all zenith angles improves the 
sensitivity to sterile neutrino mixing by about 35$\%$ over the result obtained 
using only down-going events. The improvement mainly stems from the impact of 
Earth matter effects on active-sterile mixing. The expected precision of ICAL on 
the active-sterile mixing is explored and allowed confidence level (C.L.) 
contours presented. At the assumed true value of $10^\circ$ for the sterile 
mixing angles and marginalization over $\Delta m^2_{41}$ and the sterile mixing 
angles, the upper bound at 90\% C.L. (from 2 parameter plots) is around 
$20^\circ$ for $\theta_{14}$ and $\theta_{34}$, and about  $12^\circ$ for 
$\theta_{24}$. 
}
\authorrunning{S. P. Behera et al.}
\maketitle

\section{INTRODUCTION}                                               %
A series of measurements using neutrinos from different sources {\it viz}. solar 
\cite{solar}, atmospheric, \cite{atm} reactor 
\cite{reactor1,reactor2,reactor3,reactor4}, and accelerator 
\cite{accelerator1,accelerator2,accelerator3}, have established the phenomenon 
of neutrino oscillations. The discovery of neutrino oscillations represents today major 
experimental evidence of new physics beyond the standard model. Results from these experiments led us to the current standard three-neutrino mixing paradigm, in which the three active neutrinos $\nu_{e}$, 
$\nu_{\mu}$, $\nu_{\tau}$ are superposition of three massive neutrinos 
$\nu_{1}$, $\nu_{2}$, $\nu_{3}$ with masses $m_{1}$, $m_{2}$ and $m_{3}$, 
respectively. The experimental results given by solar neutrino oscillations 
correspond to $\Delta m^{2}_{21}\simeq$ 7.5 $\times$ 10$^{-5}$ eV$^{2}$ 
and atmospheric neutrino oscillations correspond to $\Delta m^{2}_{31}\simeq 
2.4$~$\times$ 10$^{-3}$ eV$^{2}$, where $\Delta m^2_{ij}=m^2_i-m^2_j$ and $i>j$ 
with  $i,j=$1, 2, 3. Two mixing angles, which are important in the solar 
neutrino and atmospheric neutrino sectors, have been measured to be large 
($\sin^2\theta_{12}\simeq 0.3$ and $\sin^2\theta_{23}\simeq 0.5$, respectively) 
while the third mixing angle which connects the two sectors has been recently 
measured $\sin^2\theta_{13}\simeq 0.022$~\cite{global1}. 

While data from all of the above mentioned experiments fit nicely into the 
standard three generation picture, there are indications (sometimes referred as 
anomalies) from other neutrino experiments which provide a motivation to extend 
the three generation paradigm to include a fourth neutrino mixed with the three 
standard neutrinos. The first such indication comes from the LSND 
experiment~\cite{Aguilar:2001ty}, which showed an excess of electron 
anti-neutrino events above the expected background with a 3.8$\sigma$  
significance, giving a hint for $\bar\nu_{\mu} \rightarrow \bar\nu_{e}$ 
oscillations. However, the $\Delta m^2$ required to explain the data through 
neutrino oscillations is $\sim$ 1 eV$^2$, making it impossible to fit LSND data 
along with the solar and atmospheric neutrino data within the three-generation 
framework. Therefore, in order to explain the LSND results in terms of neutrino 
oscillations one has to postulate the existence of a fourth (or more) neutrino 
state(s) which is referred to as ``3~$+$~N" model, where `3' stands for active 
flavors neutrinos and `N' for sterile neutrino~\cite{Akhmedov:2010vy,Giunti:2011gz,Karagiorgi:2009nb}.
 On the other hand, from the $e^+e^-$ collider 
searches, it has been measured the number of light neutrinos coupled to be $2.92 
\pm 0.05$ \cite{pdg}. Hence, the additional light neutrino(s) may not couple to 
the standard model particles through weak currents. They are therefore referred 
to as sterile neutrinos. 

Attempts have been made at confirming or refuting the LSND hint for sterile 
neutrino oscillations. The KARMEN experiment has earlier looked for  
$\bar\nu_{\mu} \rightarrow \bar\nu_{e}$ oscillations \cite{karmen} and no 
excess observed in their data sample, conflicting with the LSND claim. However, 
due to background issues, KARMEN was unable to rule out the entire LSND region. 
The MiniBooNE experiment~\cite{min} was subsequently built to check the LSND 
signal but with baseline $L$ and energy $E$ of the experiment changed so that 
the $L/E$ matched with that of the LSND. The MiniBooNE experiment looked for 
signatures of $\nu_{\mu} \rightarrow \nu_{e}$ and $\bar\nu_{\mu} \rightarrow 
\bar\nu_{e}$ oscillations. In the $\nu_{\mu} \rightarrow \nu_{e}$ study, the 
MiniBooNE found no evidence for an excess of $\nu_{e}$ candidate events above 
475 MeV; however, a 3$\sigma$ excess of electron like events was observed below 
475 MeV, which so far has not been be explained convincingly by any known 
physics. On the other hand, in their anti-neutrino run, the 
MiniBooNE~\cite{mini2} did see an excess of $\bar\nu_e$ indicating 
$\bar\nu_{\mu} \rightarrow \bar\nu_{e}$ oscillations with $\Delta m^{2}$ 
$\simeq$ 0.1 to 1.0 eV$^{2}$  range, consistent with the evidence for 
anti-neutrino oscillations from the LSND experiment~\cite{Aguilar:2001ty}.

In addition to the long-standing LSND results, recently anomalies have emerged 
in reactor antineutrino \cite{rec,PH,Ha,Dw} and gallium-based solar neutrino 
experiments the GALLEX and the SAGE \cite{gal1,gal2}. The reactor antineutrino 
flux calculations were revisited and resulted in a decrease in the ratio of 
observed to predicted event rate from 0.976~$\pm$~0.024 to 0.943~$\pm$~0.023 
leading to deviation from unity at a 98.6~$\%$ confidence level 
(C.L.)~\cite{rec}. This in turn meant that all the earlier reactor antineutrino 
experiments has seen a deficit compared to expectation. Hence, $\Delta m^2 \sim 
1$ eV$^2$ driven active-sterile oscillations were put forth as an explanation. 
The solar neutrino experiments SAGE and GALLEX, during their calibration process 
with a $^{51}$Cr source of known strength, observed an event rate which was 
somewhat lower than expected. This again hinted towards oscillations on short 
distance scales, active-sterile neutrino oscillation with $\Delta m^2 \sim 1$ 
eV$^2$ has been proposed as an explanation. We refer the reader to 
(SAGE)~\cite{vg} for a recent status review of these anomalies (other references 
including (Reactor and Gallium) \cite{tl1}). 
 
Cosmological measurements can also provide information on the existence of 
sterile neutrinos. The sterile neutrinos having non-zero mass could impact the 
cosmic microwave background (CMB) power spectrum and modify large scale 
structure formation. The existence of sterile neutrinos which have been 
thermalized in the early Universe may contribute to the number of relativistic 
degrees of freedom (effective number of neutrino species). The combined 
analysis of data from CMB~$+$~lensing~$+$~BAO
(baryon acoustic oscillation) 
experiments~\cite{cmb} provides a robust frequentist upper limit $\sum m_{\nu}$ 
$\leq$ 0.26 eV with 95~$\%$ C.L. 

This paper presents the results of an investigation where the muon neutrino flux 
is used to constrain a possible mixing of a single sterile neutrino with the 3 
known neutrinos, $viz.$ the (3~$+$~1) model. At a high value of $\Delta m^2_{41}$ where $\Delta m^2_{41}=m^2_4-m^2_1$,
 the oscillation probability is averaged out
due to the phase part of oscillation probability. Hence, for simplicity, we have considered 
only the case of ``normal" mass hierarchy both for the active and sterile 
neutrinos.  In particular, the study quantifies the sensitivity of upcoming 
atmospheric neutrino based magnetized Iron CALorimeter detector (ICAL) at 
the India based Neutrino Observatory (INO) in constraining the active-sterile 
mixing parameters. The presence of sterile neutrinos with $\Delta m^2 \sim 1$ 
eV$^2$ leads to fast oscillations causing the suppression of the down-going 
atmospheric neutrinos, otherwise absent in the standard three-generation 
paradigm. In Ref.~\cite{raj} the authors studied the constraints on the 
active-sterile mixing expected from atmospheric neutrinos in liquid argon 
detectors and magnetized iron calorimeters. They carried out their analysis  in 
terms of neutrino energy and zenith angle, assuming fixed values of detector 
resolutions and efficiencies. In this work, a similar but a more extensive study 
is carried out. The NUANCE~\cite{nu} event generator with the ICAL detector 
geometry has been used to simulate the event spectrum at ICAL. These are then 
folded with the detector resolution functions and efficiencies obtained by the 
INO collaboration using the GEANT4 based detector simulation framework developed 
for ICAL~\cite{ac,mo}. Here, we carried out the analysis in three different 
ways. First, we binned the data in muon energy and muon zenith angle. Next we 
took into account the hadron energy information along with the muon energy and 
muon zenith angle. We also compared the results from these two set of analyses. 
Finally, we considered the neutrino events coming from all zenith angles in 
contrast to only down-going neutrino events as used in Ref.~\cite{raj} and took 
the Earth matter effects into account as well.

The outline of the paper is as follows. In Sec.~\ref{sec:icaldet} we discuss the ICAL 
detector and its physics goals. The sterile neutrino oscillation formalism 
including the Earth matter effect is introduced in Sec.~\ref{sec:osci3p1}. The 
incorporation of detector resolutions using the Monte Carlo method on neutrino 
induced raw events is given in Sec.~\ref{sec:simutech}. The procedure adopted for estimating 
the oscillated events and the binning scheme considered for estimating the 
$\chi^{2}$ using muon energy, its zenith angle and hadron energy are briefly 
explained in the same section. The definition of $\chi^{2}$ with pull is given 
in Sec.~\ref{sec:chi2esti}. The sensitivity to sterile neutrino mixing at an exposure of 1 Mt-yr 
is discussed in Sec.~\ref{sec:excllimits}. The detailed results are summarized in Sec.~\ref{sec:summ}. 
\section{\label{sec:icaldet}{INO ICAL DETECTOR}}                                                    %
The 51 kton magnetized ICAL detector, which will use iron as a target, will be 
placed in an underground cavern, with a minimum all round rock cover of 1 km, 
in order to reduce the cosmic ray background. The shape and dimensions of the 
ICAL detector have been decided keeping in mind the cavern dimensions of 132 m 
$\times$ 26 m $\times$ 32 m. The ICAL has a rectangular shape with dimensions of 
48 m $\times$ 16 m $\times$ 14.5 m. It consists of three modules each weighing 
$\sim$ 17 kton. The baseline of the ICAL magnet configuration for each of the 
three modules consists of 151 layers of low carbon steel. The layers are 
alternated with gaps of 40 mm wherein will be placed active detectors, the 
Resistive Plate Chambers (RPC), to detect the charged particles produced in 
neutrino interactions with the iron nuclei. The ICAL RPC detectors give X, Y hit 
information at $\sim$ 0.96 cm spatial resolution, the layer number gives 
Z $-$ position and timing information with a resolution ($\sigma_{t}$) $\sim$ 1 
ns~\cite{gm}. The calorimeter will be magnetized with a piecewise uniform 
magnetic field (B = 1$-$1.5 T)~\cite{spb} thereby being able to distinguish 
between the oppositely charged particles, $\mu^{-}$ and $\mu^{+}$ (produced due 
to charged current (CC) interaction of $\nu_{\mu}$ and $\bar\nu_{\mu}$, 
respectively), from the curvature of their tracks in the presence of the 
magnetic field.

The ICAL detector is mainly sensitive to muons and hence to the interaction of the
 $\nu_{\mu}$/$\bar\nu_{\mu}$. For the electron type neutrino, the detection
capability is limited because of the large iron plate thickness (5.6 cm) compared
 to the radiation length of iron ($\sim$ 1.76 cm). The production of the tau lepton
 due to the tau-neutrino interaction is also small due to high threshold for tau 
production (about 4 GeV). Due to this, the magnetized ICAL detector is most 
suited to measure muon neutrinos through the tracking of the associated muons 
and hadrons and reconstruction of their energy and momentum.
 
The GEANT4 based simulation has been carried out to study the ICAL detector 
response for the muons and hadrons. At the relevant energies, since the muon is 
a minimum ionizing particle, a clear track is produced in the ICAL detector. The 
ICAL simulation program uses the Kalman filter technique, developed by the INO 
collaboration, to reconstruct the muon track. The typical efficiency of the 
detector for a 5 GeV muon traveling vertically is about 80$\%$, while the 
typical charge identification efficiency is more than 95$\%$~\cite{ac}. The 
energy of such a muon can typically be reconstructed with an accuracy of about 
10$\%$, while its angular resolution is better than 1$^{\circ}$~\cite{ac}. The 
energy of the neutrino is the sum of the hadron (E$_h$) and muon (E$_{\mu}$) 
energies for a CC interaction. The hit multiplicity of charged particles 
distinct from the muon track can be used to estimate the total energy of hadrons 
in an event. The difference in energies of the interacting neutrino and the 
outgoing muon, E$_{h}$ $\equiv$  E$_{\nu}$ $-$ E$_{\mu}$, has been calibrated 
against the number of hits in the detector due to the shower produced by 
hadrons. The measured number of hits can then be used to reconstruct the 
fractional energy carried by the hadron from the incoming neutrino. From the 
simulation study, it has been found that the energy resolution is about 85$\%$ 
and 36$\%$ for the hadron energies of 1 GeV and 15 GeV, respectively, in the 
central region of the ICAL detector \cite{mo}. Although the energy resolution of 
hadrons is much lower than that for muons, it still gives an additional 
information about the particular event, which can be used to improve the
physics reach of the detector \cite{mon,dal}.

The physics goal of the ICAL detector is to measure precisely the atmospheric
neutrino mixing parameters, {\it viz}. $\Delta m_{31}^{2}$ and 
sin$^{2}2\theta_{23}$, \cite{precision} and in particular, determine the 
neutrino mass hierarchy by measuring the sign of $\Delta m_{31}^{2}$ 
\cite{mon,mh}. In addition, it can be used for octant sensitivity study $i.e.$ 
whether $\theta_{23}$ is maximal or not, and if it is indeed non-maximal, 
whether $\theta_{23}$ is less than 45$^\circ$ or greater than 
45$^\circ$~\cite{dal}. The ICAL experiment would also be able to put severe 
constraints on new physics scenarios like CPT violation~\cite{cot}.
 
\section{\label{sec:osci3p1}{Oscillation probability using 3~$+$~1 model}}              %
The sterile neutrino oscillation probabilities are based on expansion of the 3 
generation Pontecorvo-Maki-Nakagawa-Sakata (PMNS) matrix to 3~$+$~1 generation, 
where ``3" stands for active and ``1" for sterile neutrino, respectively. The 
neutrino flavors and mass eigenstates are related through
\begin{equation}
\begin{pmatrix}
\nu_{e}\\
\nu_{\mu}\\
\nu_{\tau}\\
\nu_{s}
\end{pmatrix}
=
\begin{pmatrix}
U_{e1} & U_{e2} & U_{e3} & U_{e4} \\
U_{\mu 1} & U_{\mu 2} & U_{\mu 3} & U_{\mu 4} \\
U_{\tau 1} & U_{\tau 2} & U_{\tau 3} & U_{\tau 4} \\
U_{s1} & U_{s2} & U_{s3} & U_{s4} \\
\end{pmatrix}
\begin{pmatrix}
\nu_{1}\\
\nu_{2}\\
\nu_{3}\\
\nu_{4}
\end{pmatrix}
\label{eq:mixingmatrix}\textrm,
\end{equation}
where $U$ is the mixing matrix. In this analysis the following parametrization 
has  been considered
\begin{equation}
U = R(\theta_{34})R(\theta_{24})R(\theta_{23})R(\theta_{14})R(\theta_{13})R(\theta_{12}),
\end{equation} 
where $R(\theta_{ij})$ are the (complex) rotation matrices and $\theta_{ij}$ are 
the mixing angles with $i,j$ = 1, 2, 3, 4; and the order of rotation angles 
are considered from Ref.~\cite{Maltoni:2007zf}. Using the above definition, 
neutrino flavor change can be described as a function of the mixing matrix 
elements and neutrino masses in terms of the neutrino oscillation probability
\begin{equation}
\begin{split}
P_{\alpha\beta} = \ & \delta_{\alpha\beta}-4\sum_{i>j}Re(U_{\alpha i}U^{*}_{\beta i}U^{*}_{\alpha j}U_{\beta j})\sin^{2}\frac{\Delta m^{2}_{ij}L}{4E}\\ & + 2 \sum_{i>j}Im(U_{\alpha i}U^{*}_{\beta i}U^{*}_{\alpha j}U_{\beta j})\sin^{2}\frac{\Delta m^{2}_{ij}L}{2E},
\end{split}
\label{eq:prob}
\end{equation}
where $\alpha$, $\beta$ = e, $\mu$, $\tau$, s; $\Delta 
m^{2}_{ij}$= $m^{2}_{i}- m^{2}_{j}$ with $i>j$, and $L$ is the source to 
detector distance and $E$ is the energy of neutrinos. The order of rotation 
angles shows that, if all mixing angles are zero, then there is a correspondence 
between the flavor and mass basis $i.e.$ ($\nu_{e}$, $\nu_{\mu}$, 
$\nu_{\tau}$, $\nu_{s}$) = ($\nu_{1}$, $\nu_{2}$, $\nu_{3}$, $\nu_{4}$). 
Assuming  zero CP phase in lepton sector, in total, there are four new mixing 
parameters are introduced in the (3~$+$~1) neutrino model. The three mixing 
angles ($\theta_{14}$, $\theta_{24}$, $\theta_{34}$),  and one new mass-squared 
difference, which we choose to be  $\Delta m^{2}_{41}$= $m^{2}_{4}- m^{2}_{1}$. 
It is to be noted that other two extra mass-squared differences, $\Delta 
m^{2}_{42}$ and $\Delta m^{2}_{43}$, are not independent and can be expressed 
as $\Delta m^{2}_{42}$ =  $\Delta m^{2}_{41} - \Delta m^{2}_{21}$ and $\Delta 
m^{2}_{43}$ =  $\Delta m^{2}_{41} - \Delta m^{2}_{31}$. The global analysis on 
the sterile neutrino mixing has been carried out in Ref.~\cite{Kopp:2013vaa}.
 The best fit values of the parameters $\lvert U_{e4}\rvert^{2}$, 
$\lvert U_{\mu4}\rvert^{2}$ and $\Delta m^{2}_{41}$, characterizing the 
active-sterile neutrino (antineutrino) mixing in the 3~$+$~1 scheme are,
\begin{equation*}
\lvert U_{e4}\rvert^{2} = 0.0225,\, \lvert U_{\mu4}\rvert^{2} = 0.0289,\, 
\Delta m^{2}_{41} = 0.93 \,\,
\textrm{eV}{^2}, 
\end{equation*}   
where $\Delta m^{2}_{41}$ = $m^{2}_{4} - m^{2}_{1}$. 
\begin{figure*}
\includegraphics[width=0.85\textwidth]{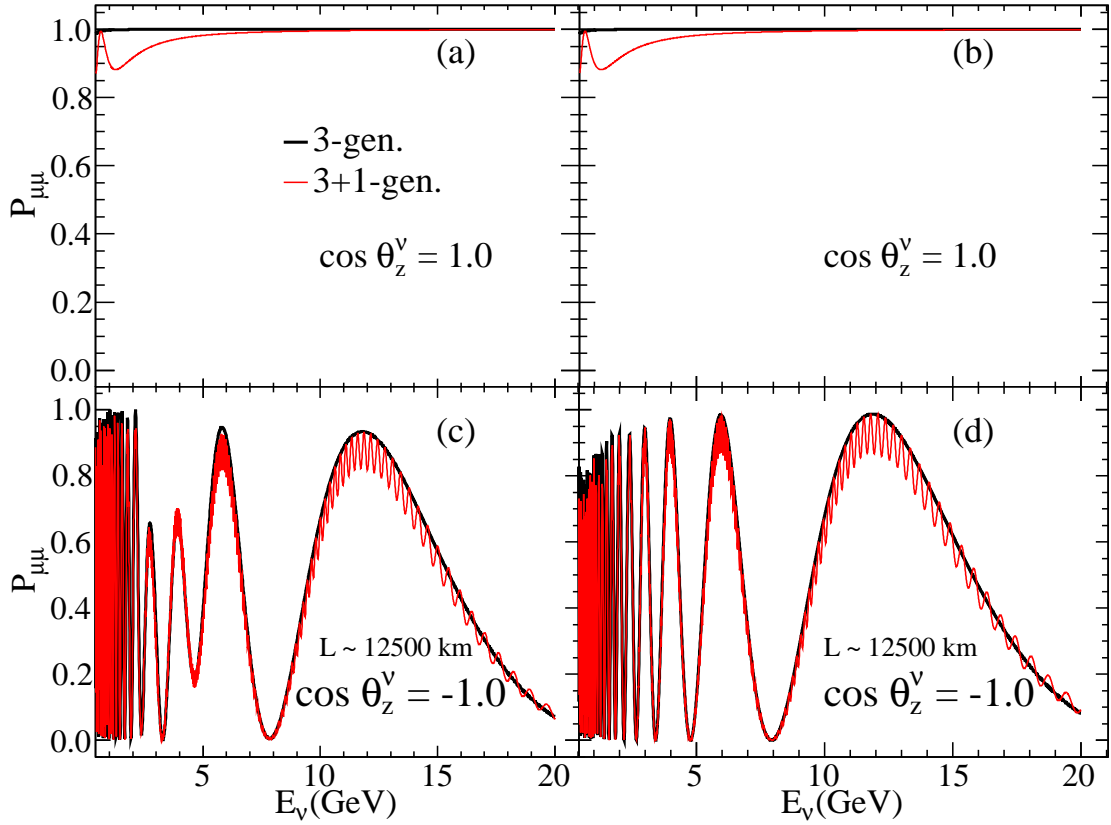}
\caption{Comparison of the survival probability for neutrinos 
(left-hand panels) and anti-neutrinos (right-hand panels) for 3 (black line) and 
3~$+$~1 (red line) generations, as a function of (anti)neutrino energy. The top 
panels are for $\cos\theta^{\nu}_{z} =1.0$ (down-going neutrinos, L = 15 km) while the 
bottom panels are for $\cos\theta^{\nu}_{z} =-1.0$ (up-coming neutrinos, L = 12500 km). The 
oscillation parameters assumed for all panels are $\theta_{24} = 10.0^{\circ}$ , 
$\theta_{14}= \theta_{34} = 0.0^{\circ}$, $\Delta m^{2}_{41} = 0.1$ eV$^{2}$, 
$\Delta m_{31}^2 = 2.4 \times 10^{-3}$ eV$^2$, $\Delta m_{21}^2 = 7.5\times 
10^{-5}$ eV$^2$, $\sin^{2}\theta_{23}=0.5$, $\sin^{2} \theta_{12}=0.3$, 
$\sin^{2} 2\theta_{13}=0.1$ and all CP phases are taken as zero.}
\label{fig:prob}
\end{figure*}
\subsection{Matter effect on sterile neutrino oscillation}               %
 When atmospheric neutrinos travel through matter, 
they undergo coherent forward scattering on matter. The scattering is mediated 
by both charged and neutral current processes. The effective mass matrix in matter 
then changes to 
\begin{equation}
M^{2}_{F} = U M^{(3~+~1)} U^{\dagger} + A\text,
\end{equation}
where
\begin{equation}
M^{(3~+~1)} = diag(m^{2}_{1},m^{2}_{2},m^{2}_{3},m^{2}_{4}),
\end{equation}
\begin{equation}
A = diag(A_{CC},0,0,A_{NC}),
\end{equation}
where
\begin{equation}
A_{CC} = \pm 2\sqrt2 G_{F}\rho N_{A} Y_{e} E\text,
\label{eq:acc}
\end{equation}
and 
\begin{equation}
A_{NC} = \pm \sqrt2 G_{F}\rho N_{A} (1-Y_{e}) E .
\label{eq:anc}
\end{equation} 
Here, $A_{CC}$ and $A_{NC}$ are the matter induced weak charged current (CC) and 
neutral current (NC) potentials, respectively, which depend on Fermi's constant, 
$G_{F}$, matter density $\rho$, Avogadro number $N_{A}$, electron fraction 
$Y_{e}$ in matter and energy of neutrino $E$. In Eqs.~(\ref{eq:acc}) and 
(\ref{eq:anc}), the ``+" and ``$-$" sign corresponds to neutrinos and 
antineutrinos, respectively. Only $\nu_{e}$ has CC interaction with electrons and 
NC interaction with electrons and nucleons, whereas $\nu_{\mu}$  and 
$\nu_{\tau}$ have only NC interaction and sterile neutrinos have no weak 
interactions. The analysis including matter effect has been carried out 
considering the varying density profile of the Earth $i.e.$ the preliminary 
reference Earth model (PREM) \cite{pre}. Due to the matter effect, oscillation 
probability, as given in Eq.~\ref{eq:prob} will be modified, and in this work 
the neutrino oscillation probabilities are numerically calculated in matter for 
exact 3~$+$~1 generation framework. 

Figure~\ref{fig:prob} shows the survival probabilities for neutrinos (left-hand 
panels) and anti-neutrinos (right-hand panels) for 3 and 3~$+$~1 generations, as 
a function of (anti)neutrino energy. The values of the parameters chosen to 
obtain the oscillation probabilities are given in the figure caption. The black 
and red  lines show the plot for 3 and 3~$+$~1 generation neutrinos, 
respectively  considering the Earth matter effect. The top panels are for 
$\cos\theta^{\nu}_{z} =1.0$ (down-going neutrinos) while the bottom panels are 
for $\cos\theta^{\nu}_{z} =-1.0$ (upcoming neutrinos), where $\theta^{\nu}_{z}$ 
is the neutrino zenith angle. We show here the probabilities for the case when 
only the $\theta_{24}$ sterile mixing angle is taken as non-zero while 
$\theta_{14}=0=\theta_{34}$. The values of the other oscillation parameters are 
taken close to their current best-fit values and are given explicitly in the 
figure caption. The top panels show that while there is no oscillation of 
neutrinos for the standard 3 generation case whereas the 3~$+$~1 generations 
show a small depletion which is same for neutrinos and antineutrinos. This is 
due to all CP phases being zero and oscillations happening in vacuum. The 
depletion which comes from active-sterile mixing can be used to constrain 
the sterile neutrino framework \cite{raj}. The bottom panels show the presence 
of additional Earth matter effects due to active-sterile mixing. The rapid 
oscillation for 3~$+$~1 generation neutrinos is due to the $|\Delta 
m^2_{41}|-$ driven phase factor  of the oscillation probability. The ICAL 
detector is not expected to decipher these fast oscillations. For this 
$\cos\theta_z^{\nu}$ the neutrino and antineutrino survival probabilities have 
a difference coming mainly from $\Delta m_{31}^2-$ driven earth matter effects, 
while the impact due to sign of $\Delta m_{41}^2$ is hardly evident for this 
case where the mixing angles $\theta_{14}$ and $\theta_{34}$ are taken as zero. 

\begin{figure*}[t]
\centering
\includegraphics[width=0.85\textwidth]{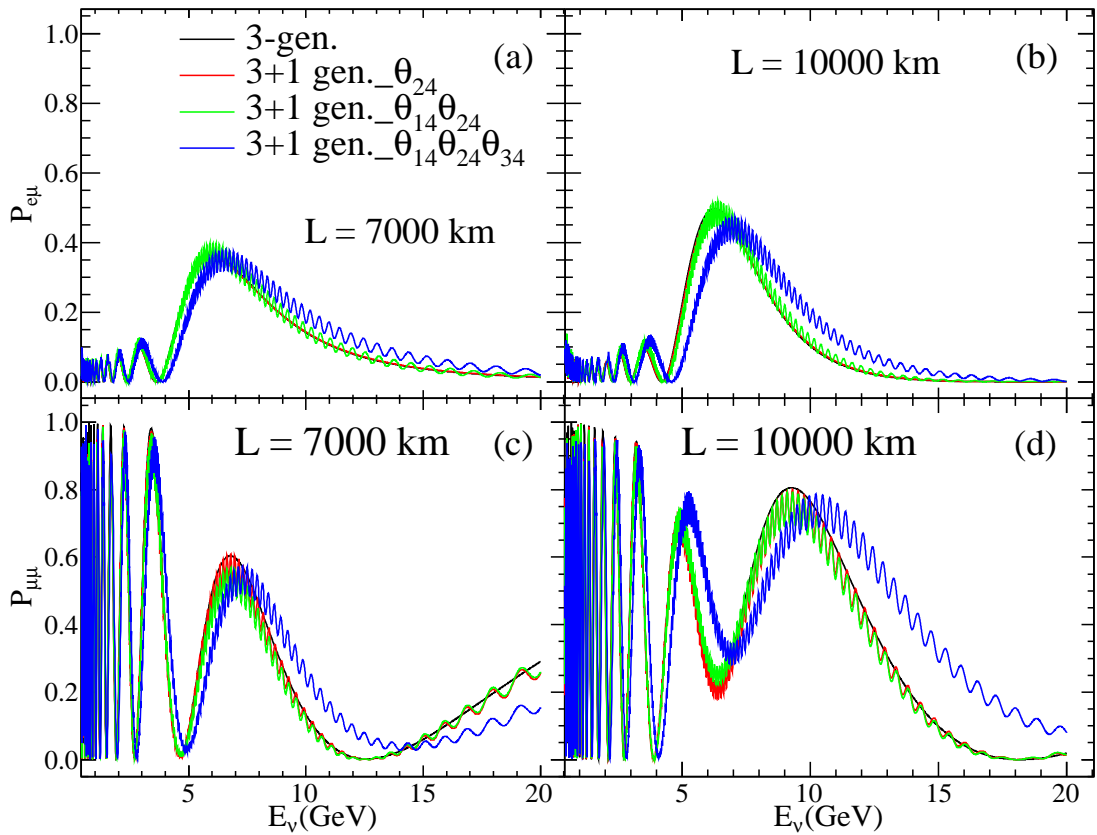}
\caption{\label{fig:pcomb}{The neutrino oscillation probability P$_{e\mu}$ 
(top panels) and P$_{\mu\mu}$ (bottom panels) as a function of neutrino energy. 
The left-hand panels (right-hand panels) are for a neutrino trajectory 
corresponding to a baseline of $L=7000$ km ($L=10000$ km) inside earth. 
The black lines correspond to the 3 generation case, while the red, green,
and blue lines correspond to the 3~$+$~1 scenario. Specifically, the red lines 
are for $\sin^2\theta_{24}=0.03$, $\sin^2\theta_{14}=0$ and 
$\sin^2\theta_{34}=0$, the green lines are for $\sin^2\theta_{24}=0.03$, 
$\sin^2\theta_{14}=0.022$ and $\sin^2\theta_{34}=0$, while the blue lines are 
for $\sin^2\theta_{24}=0.03$, $\sin^2\theta_{14}=0.022$ and 
$\sin^2\theta_{34}=0.21$. The other oscillation parameters assumed for all 
panels are $\Delta m^{2}_{41} = 1.0 $ eV$^{2}$, $\Delta m_{31}^2 = 2.4 \times 
10^{-3}$ eV$^2$, $\Delta m_{21}^2 = 7.5\times 10^{-5}$ eV$^2$, 
sin$^{2}\theta_{23}=0.5$, $\sin^{2} \theta_{12}=0.3$, $\sin^{2} 
2\theta_{13}=0.1$  and all CP phases are taken as zero. }
}
\end{figure*}
\par Fig.~\ref{fig:pcomb} shows the appearance P$_{e\mu}$ (top panels) and 
survival, P$_{\mu\mu}$ (bottom panels), neutrino oscillation probabilities for 3 
and 3~$+$~1 generations, as a function of neutrino energy. The values of the 
mixing parameters chosen to obtain the oscillation probabilities are given in 
the figure caption.  In particular, one can see that here we allow the sterile 
mixing angles $\theta_{14}$ and $\theta_{34}$ to be non-zero and study the 
impact of these on the oscillation probabilities. The left-hand panels have 
been obtained for a baseline of $L=7000$ km (corresponding to a zenith angle of 
$\cos\theta_z^\nu=-0.55$) and right-hand panels are for a baseline of $L=10000$ km 
(corresponding to a zenith angle of $\cos\theta_z^\nu=-0.785$). The black lines 
show the plot for the 3 generation case and red, green and blue lines represents 
for the 3~$+$~1 generation neutrinos case. The red lines correspond to the case 
where only $\theta_{24}$ is taken as non-zero ($\sin^2\theta_{24}=0.03$ here). 
This is similar to the plots shown in Fig. 1. We next show in green lines the 
change obtained when we make $\theta_{14}$ non-zero. We notice a small 
difference between the red and green curves. Finally, in the blue lines we take 
all the 3 sterile mixing angles to be non-zero and their values are given in the 
figure caption. We see that the oscillations probabilities change significantly 
when $\theta_{34}$ is switched-on. Note that the oscillation probabilities 
P$_{e\mu}$ and P$_{\mu\mu}$ are independent of the value of $\theta_{34}$ 
as per the parametrization of the mixing matrix, and the effect of 
$\theta_{34}$ comes entirely due to earth matter effects \cite{Choubey:2007ji}. 
Both the oscillation probabilities, P$_{e\mu}$ and P$_{\mu\mu}$ also show a 
shift in position of maxima or minima with inclusion of $\theta_{34}$. 
We notice from the figure that while each of the sterile mixing parameters 
changes the neutrino oscillation probabilities, the impact of $\theta_{34}$ 
appears to be most dramatic and changes the shape of the oscillation 
probabilities. It may be noted that, in this analysis both disappearance 
($\nu_{\mu} \rightarrow \nu_{\mu}$) and appearance ($\nu_{e} \rightarrow 
\nu_{\mu}$) oscillation channels are considered while estimating the 
oscillated muon events for the atmospheric neutrinos. 
%
\begin{figure*}[t]
\centering
\includegraphics[width=0.95\linewidth,height=0.35\linewidth]{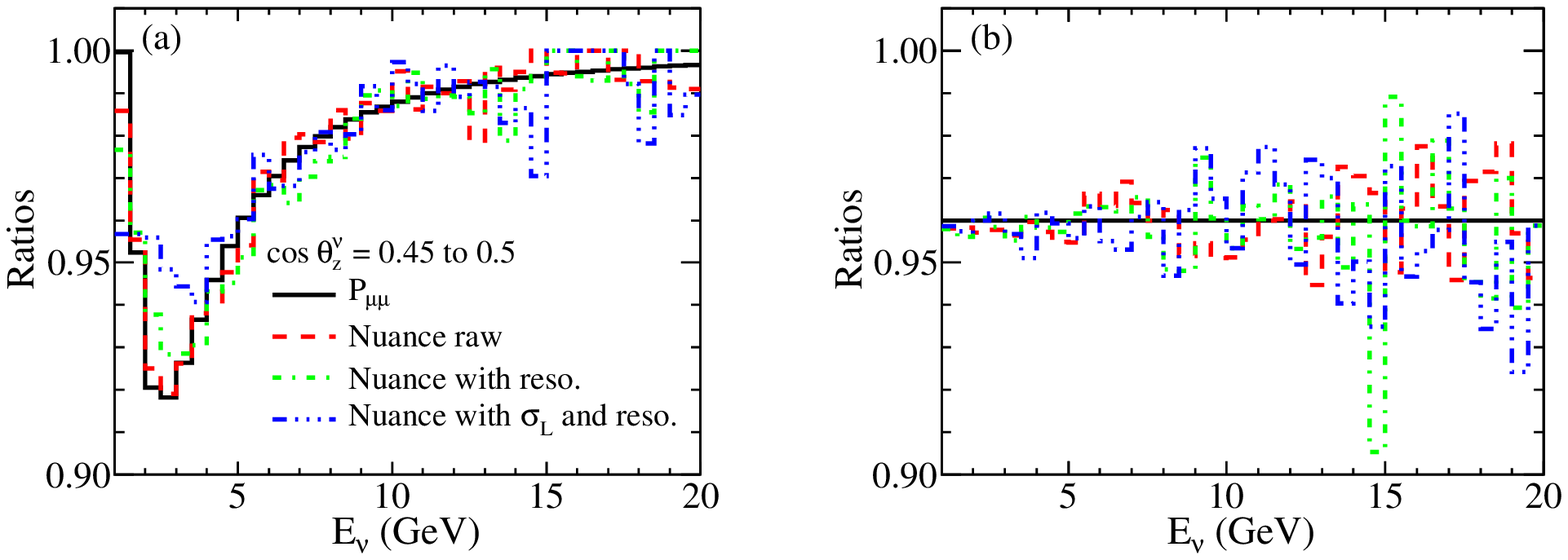}
\caption{\label{fig:ev1}{The ratio of oscillated to unoscillated events versus 
neutrino energy for the $\cos\theta_{z}$ bin 0.40 to 0.45. The active-sterile 
oscillation parameters are taken as $\theta_{14}=\theta_{34}=0$, 
$\sin^22\theta_{24}=0.083$ and (a) $\Delta m^{2}_{41}$ = 0.1 eV$^{2}$, (b) 
$\Delta m^{2}_{41}$ = 10.0 eV$^{2}$. The black solid lines show the plot using 
the neutrino oscillation formula, the red dashed lines show the plot obtained 
using the NUANCE events without implementing the detector response, the green 
dot-dashed lines shows the plot using events after implementing the detector 
response of $\sigma_{E_{\nu}}$ = 0.15 E$_{\nu}$ and $\sigma_{\theta_{z}}$ = 10$^\circ$, 
and the blue dashed-dotted lines show the plot for events after incorporating 
the detector response as well as the production height distribution of neutrinos 
given in Table I.}}
\end{figure*}
\begin{figure*}[ht]
\centering
\includegraphics[width=0.45\linewidth,height=0.4\linewidth]{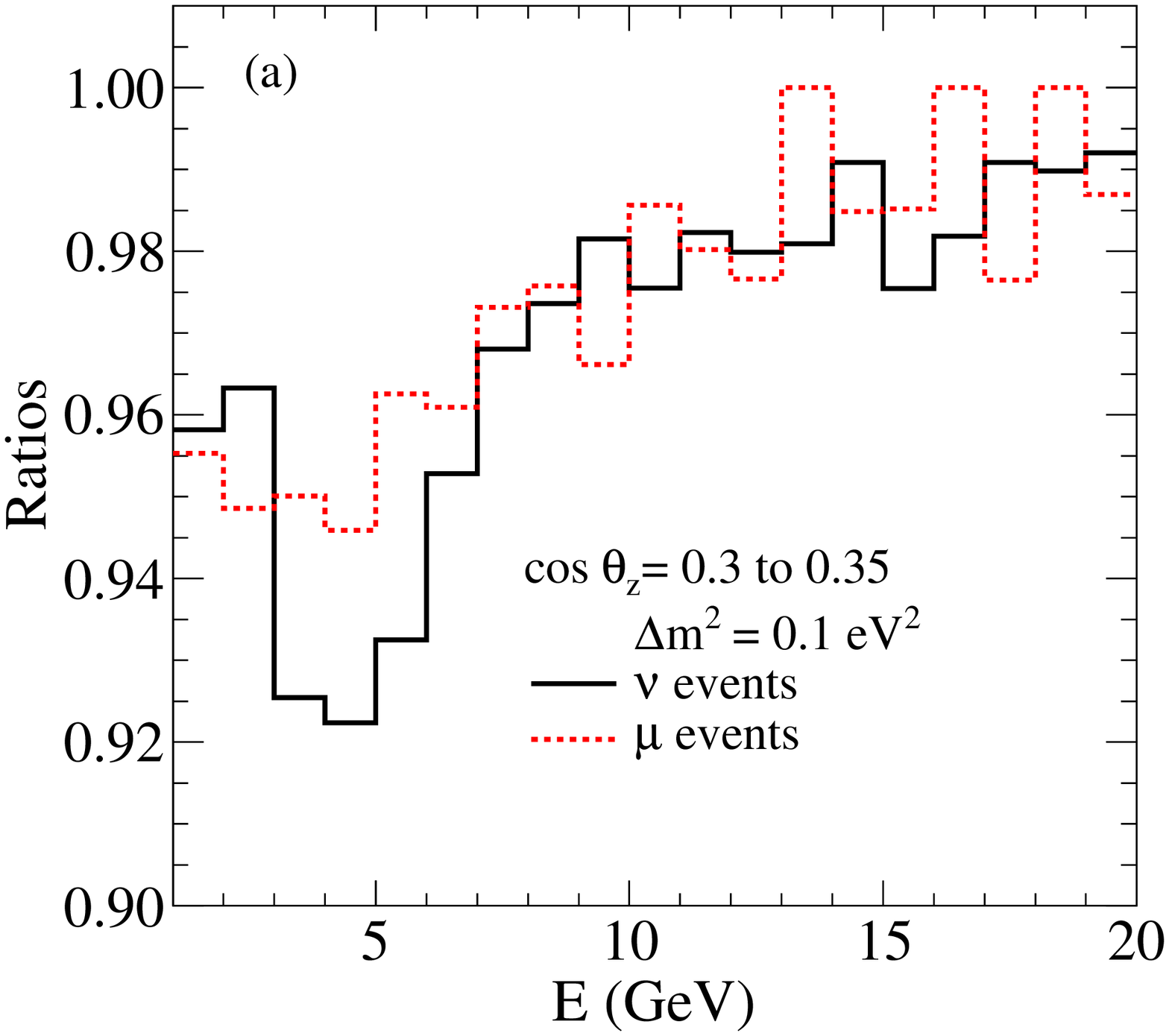}
\includegraphics[width=0.45\linewidth,height=0.4\linewidth]{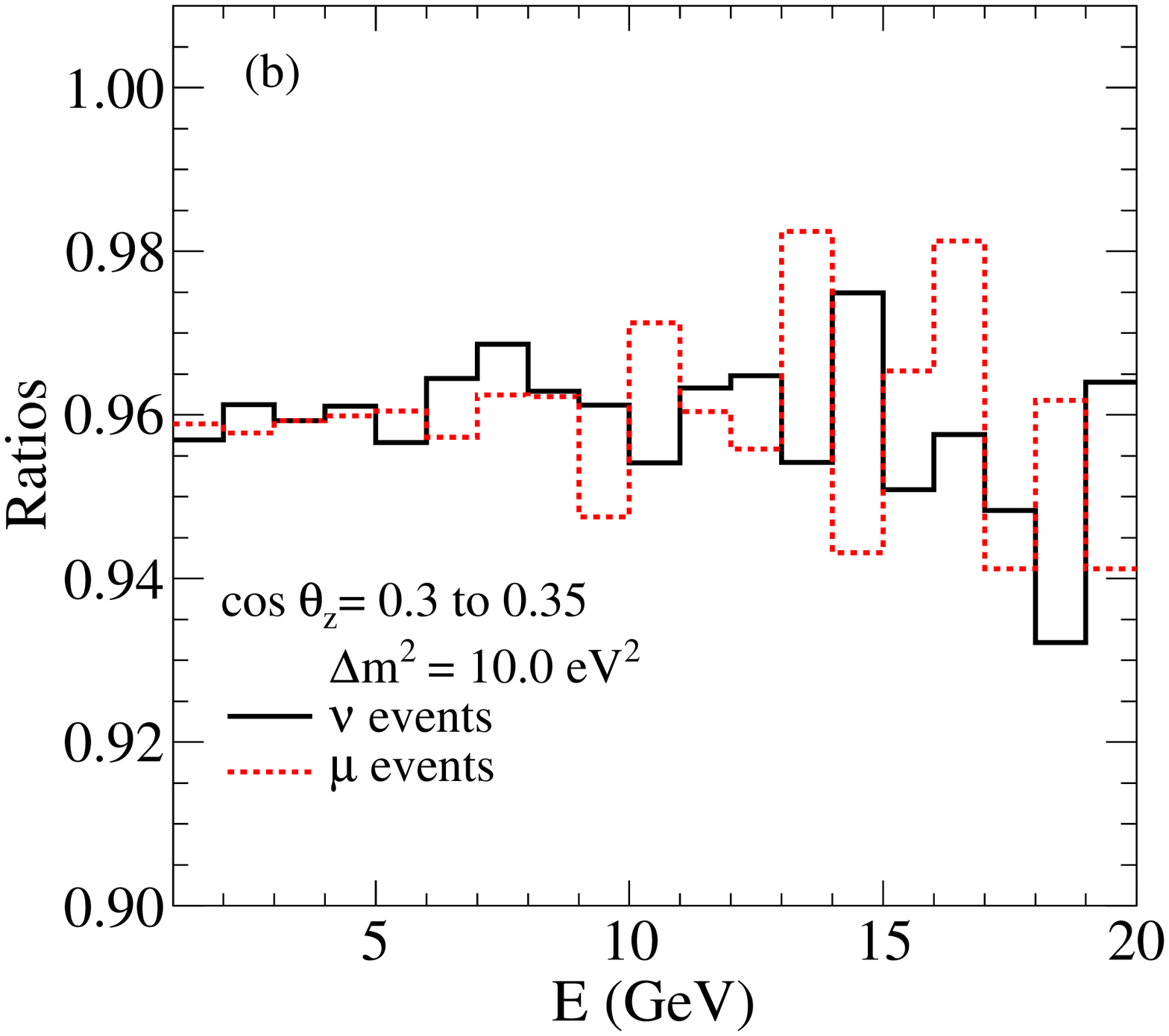}
\caption{\label{fig:ev2}The ratio of  oscillated to unoscillated 
events as a function of neutrino (black solid lines) and muon (red dashed lines) 
energy. The cos $\theta_{z}$ bin chosen is 0.30 to 0.35, 
$\theta_{14}=\theta_{34}=0$, $\sin^22\theta_{24}=0.083$ and (a) $\Delta 
m^{2}_{41}$ = 0.1 eV$^{2}$, (b) $\Delta m^{2}_{41}$ = 10.0 eV$^{2}$. The 
standard oscillation parameters are same as in Fig.~\ref{fig:ev1}.}
\end{figure*}
%
\section{\label{sec:simutech}{Simulation Technique}}%
In this section we describe the simulation of atmospheric neutrino events 
in ICAL. The generation of the event spectrum and binning scheme are presented 
in subsection~\ref{subsec:eventrec} and in subsection~\ref{subsec:binsch}, respectively. 
\subsection{\label{subsec:eventrec}{EVENT RECONSTRUCTION}}                
The study on sterile neutrino sensitivity has been carried out using the 
detailed information of neutrino induced events. The atmospheric neutrinos, 
interacting with the iron target, produce leptons and hadrons through the charge 
current interaction. There are mainly three processes which contribute to the 
CC interactions in the ICAL detector. At the sub-GeV energy range, the 
quasi-elastic process dominates, in which the final state muon carries most of 
the available energy. As the energy increases from sub-GeV to multi-GeV, hadrons 
and their showers are produced in resonance (RS) and deep-inelastic scattering 
(DIS) processes. In the RS process, most of the hadron showers consist of a 
single pion, though multiple pions may contribute in a small fraction of 
events. On the other hand, in the DIS process multiple hadrons are produced 
which carry a large fraction of the incoming neutrino energy. In the CC 
interaction of neutrino, for every hadron shower, there is a corresponding muon 
coming from the same CC interaction vertex of neutrino. 
 
Again, as shown in Fig.~\ref{fig:prob}, the probability of active-sterile 
oscillations is valid only for down-going neutrinos, if one uses vacuum 
oscillations. However, the inclusion of the Earth matter effects for the upward 
going neutrino may contribute to the active-sterile oscillations. In what 
follows, we will consider neutrino events covering all zenith angles and present 
a comparison of the improvement coming from upcoming events.  

The raw events without oscillations are generated using the NUANCE neutrino 
event generator modified for the ICAL detector. To reduce statistical 
fluctuations, initially data are generated for the 1000 years and further 
normalize to the required exposure during the statistical analysis. In this 
analysis, an exposure of 1 Mt-yr has been considered. Neutrino oscillations are 
then introduced through a re-weighting algorithm as discussed in 
\cite{Ghosh:2012px}, which is based on the acceptance-rejection method. While 
calculating the neutrino oscillation probabilities, we consider the $\nu$ 
production height distribution in the atmosphere as a function of zenith angle 
and energy of neutrino which is given in \cite{TK} for only downgoing neutrino. 
The path length (L) distribution was incorporated on MC basis by considering 
Gaussian smearing of the path length whose mean is L and the corresponding 
zenith angle and energy dependent sigma as given in Table I. It is to be noted 
that we have considered the production altitude distribution for down-going 
neutrinos only. This is because the sterile neutrino oscillations are relevant 
on small length scales and the uncertainty in the production point for the 
down-going neutrinos, therefore, becomes crucial.
%
\begin{table}[h]
\caption{Production height (slant distance in km) of neutrinos and its 
corresponding sigma for six values of cos$\theta_z$ at three neutrino energy 
ranges \cite{TK}}
\begin{tabular}{|l|c c|cc|c c|}\hline
cos$\theta_{z}$  & \multicolumn{2}{m{2.2cm}|}{\centering E = 0.3 $-$ 2.0 [GeV]} 
& \multicolumn{2}{m{2.2cm}|}{\centering E = 2.0 $-$ 20.0 [GeV]} & 
\multicolumn{2}{m{2.2cm}|}{\centering E $>$ 20.0 [GeV]}\\
\cline{2-7}
& L & $\sigma_{L}$ & L & $\sigma_{L}$ & L & $\sigma_{L}$\\
\hline
1.0  & 15.9 & 8.7  &  16.6 & 9.0  & 17.6 & 8.9 \\
0.75 & 23.6 &11.8  &  24.1 & 12.1 & 25.8 & 12.6\\
0.50 & 41.0 &18.1  &  40.9 & 19.1 & 43.3 & 19.4\\
0.25 & 95.6 &31.4  &  92.8 & 34.6 & 94.9 & 36.4\\
0.15 & 160.0&37.3  &  154.3& 42.8 & 151.2& 49.4\\
0.05 & 369.8&55.0  &  359.0& 67.1 & 335.7& 94.2\\
\hline
\end{tabular}
\end{table}
%
Figure~\ref{fig:ev1} shows the ratio of events with and without oscillations 
taken into account as a function of neutrino energy. The results are shown for 
the neutrinos and a particular zenith angle bin of $\cos\theta_z^\nu = 
0.45-0.5$. Fig.~\ref{fig:ev1}(a) is for $\Delta m_{41}^2 = 0.1$ eV$^2$ while 
Fig.~\ref{fig:ev1}(b) is for $\Delta m_{41}^2 = 10.0$ eV$^2$. The black lines 
show the survival probability $P_{\mu\mu}$ as a function of the neutrino energy. 
The red line shows the ratio of raw events from NUANCE with and without 
oscillations. Note that the red lines follow the survival probability fairly 
close, within the MC fluctuation. The green lines are obtained by including a 
flat energy and angle resolution for the neutrino events. We assume a flat 
resolution of $\sigma_{E_\nu}$~=~0.15~$E_\nu$ and 
$\sigma_{\theta_z^\nu}$~=~$10^\circ$ in this illustrative figure. It may be 
noted that, the energy and the zenith angle correspond to true values obtained 
from the NUANCE output. On the other hand, the reconstructed energy and zenith 
angle correspond to the neutrino events after incorporating detector 
resolutions. The resolution functions bring a mild smearing of the shape of the 
event spectrum. Finally, the blue lines are obtained after smearing for the 
production point in the atmosphere as well according to Table 1. The impact of 
this smearing is rather large when the data is sensitive to the phase of the 
$\Delta m_{41}^2$ driven oscillations (left panel). Since the production point 
becomes uncertain, the path length becomes uncertain, causing a drop in the 
effective dip of the event spectrum due to active sterile oscillations. 
 
The ICAL simulations are performed not in terms of the neutrino energy and 
angle, but in terms of muon energy and angle and hadron energies. Hence the 
oscillated events are  distributed two dimensionally in terms of final state 
muon energy and muon zenith angle bins. The hadron events are binned in hadron 
energy only. The detector energy resolution, angle resolution, charge and 
reconstruction efficiencies are next incorporated in the simulations. We have 
incorporated the actual resolutions of the detector for muons and hadrons using 
INO look up tables~\cite{ac,mo}. The muon energy and zenith angle resolutions 
are a function of both energy as well as zenith angle. To incorporate the 
detector response in the true event distribution numerically, we have used the 
Monte Carlo (MC) methods. At low energy (E $<$ 0.9 GeV), the energy loss of 
muons in the detector follow a Landau probability distribution function (p.d.f) 
and above this a Gaussian p.d.f. For a particular event, the Landau p.d.f 
(P$_{L}$) for E$_{\mu}$ $<$ 0.9 GeV and Gaussian p.d.f. (P$_{G}$) above 0.9 GeV 
have been used to smear the true energy of muon. To incorporate the energy 
dependent zenith angle resolution of muon, the P$_{G}$ has been used for all 
energy. The mean of the function are true $E_{\mu}$, cos $\theta_{z}$ for energy 
and zenith angle of muons, respectively, without incorporating the ICAL detector 
resolutions. The final energy and cosine of zenith angles are as follows,  
\begin{equation}
E^{r}_{\mu} = P_{L}(E_{\mu},\,\, \sigma_{E_{\mu}}), E < \,\text 0.9\, GeV
\end{equation}
\begin{equation}
E^{r}_{\mu} = P_{G}(E_{\mu},\,\, \sigma_{E_{\mu}}), E \ge\, \text 0.9\, GeV
\end{equation}
\begin{equation}
cos^{r}\theta_{z} = P_{G}(cos \ \theta_{z}, \sigma_{cos \theta_{z}})
\end{equation}
 where $\sigma$$_{E_{\mu}}$ is the standard deviation of energy, 
 $\sigma$$_{cos {\theta_z}}$ is the standard deviation for cosine of zenith 
angle taken from \cite{ac}, and $E^{r}_{\mu}$, $cos^{r}\theta_{z}$ are 
reconstructed energy and cosine of zenith angle of muons, respectively. To 
incorporate the hadron energy resolutions in the ICAL simulation, the Vavilov 
probability  distribution function has been used as described in \cite{mo}. 

\begin{figure*}[]
\centering
{\includegraphics[width=0.4\linewidth,height=0.35\linewidth]{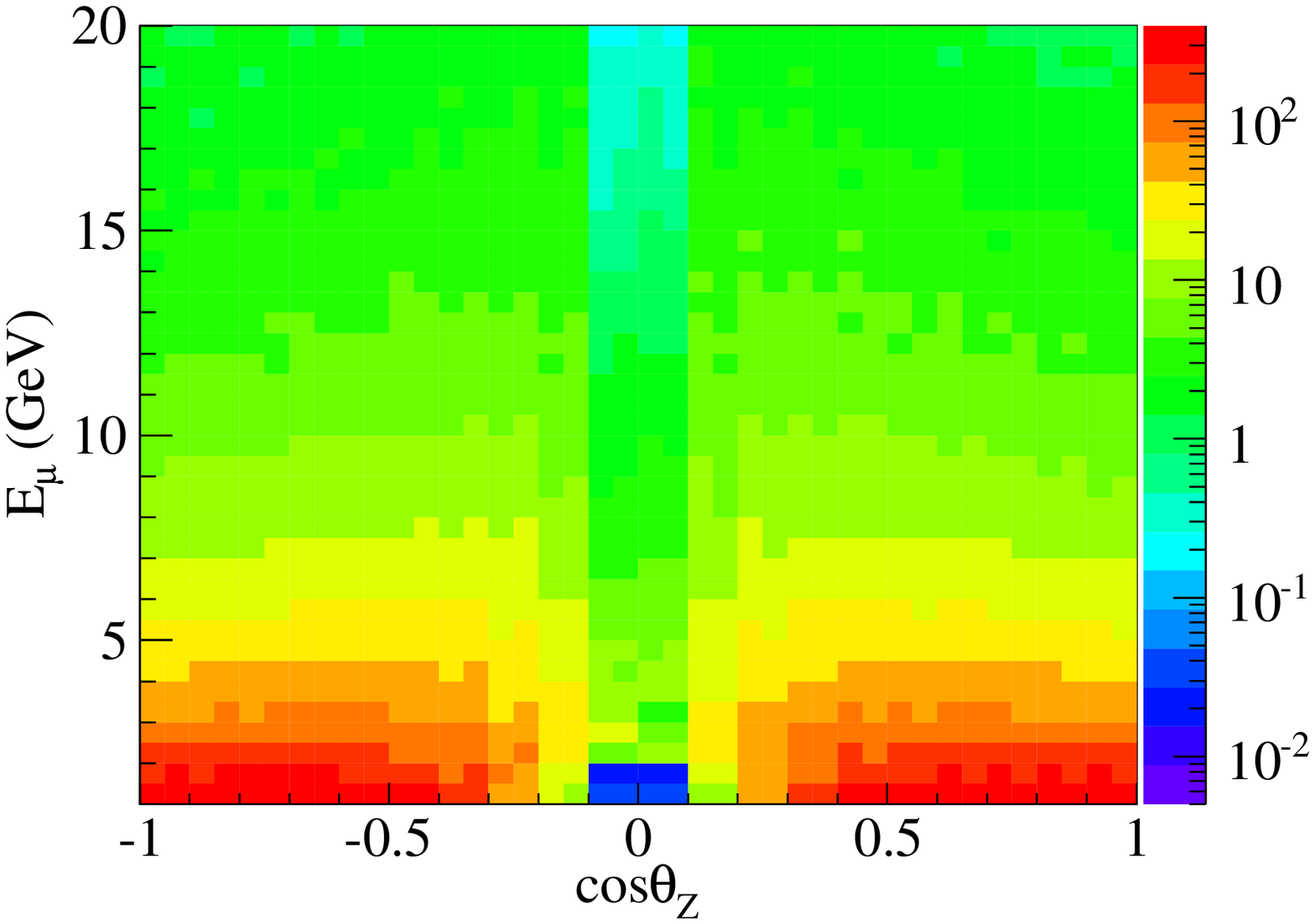}}
{\includegraphics[width=0.4\linewidth,height=0.35\linewidth]{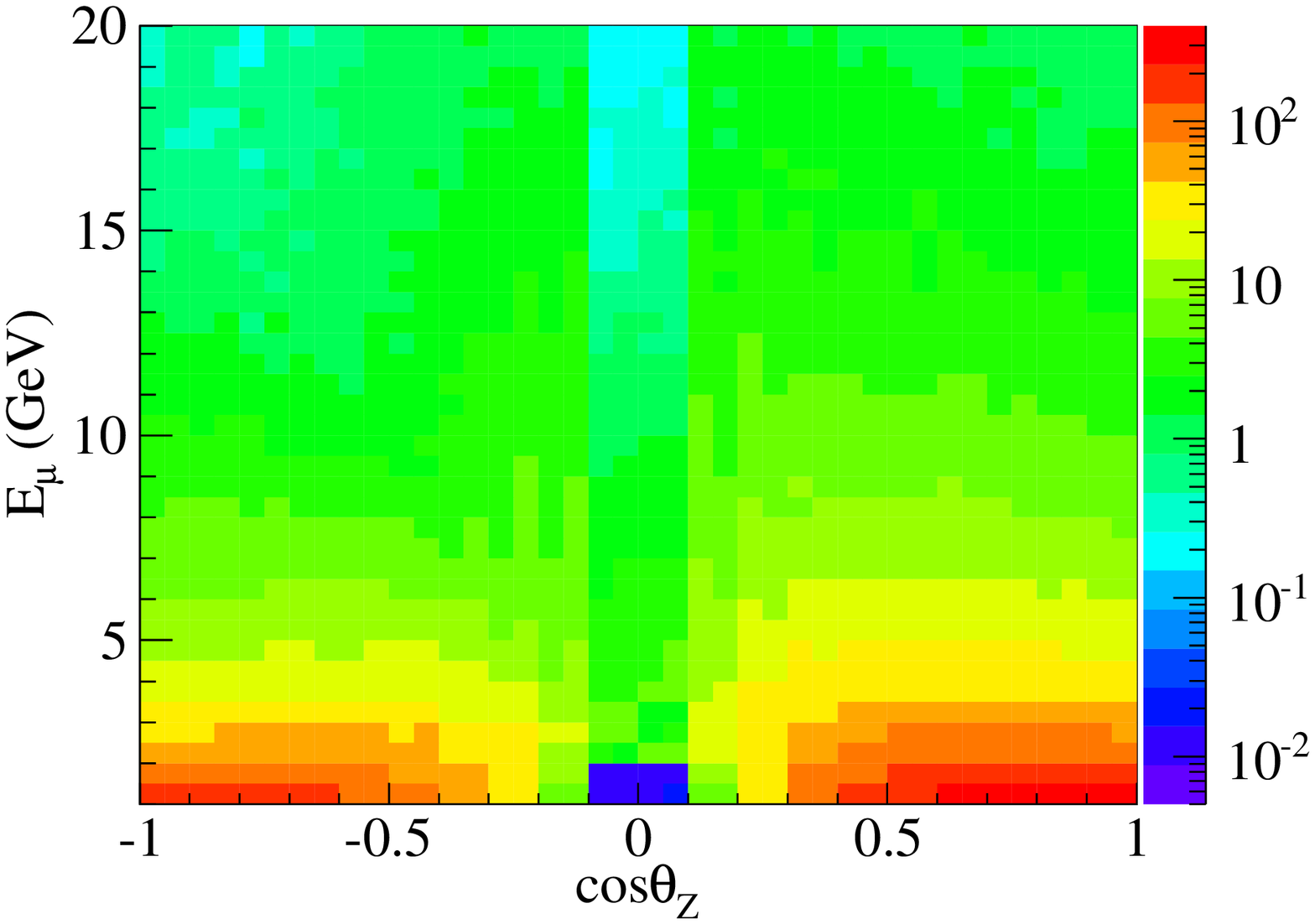}}
   \caption{\label{fig:event}(a) Without oscillated event distribution (b) Oscillated event distributions.}
    \label{fig:fig1}
    \end{figure*}

In Fig.~\ref{fig:ev2} the red dashed lines show the ratio of events, with and 
without oscillations, binned in  muon energy and in the muon zenith angle range 
 $\cos\theta_{z}$ = 0.3 to 0.35. For comparison we also show in the black solid 
lines the ratio of events, with and without oscillations, binned in terms of 
neutrino energies and in the neutrino zenith angle range of $\cos\theta_{z}^\nu$ 
= 0.3 to 0.35. The left panel (a) is for $\Delta m_{41}^2=0.1$ eV$^2$ while the 
right panel (b) is for $\Delta m_{41}^2=10.0$ eV$^2$. The number of downgoing without oscillated events are 12984
and oscillated events are 11868 for an exposure of 1 MTon-year.
The expected rate of neutrinos assuming no sterile neutrinos as a function
of cosine of zenith angle (cos$\theta_z)$ and energy (E$_{\mu}$ is shown in
Fig.~\ref{fig:event}(a). The relative reduction in 
expected neutrinos events rate for the mixing parameters around the best fit~\cite{Kopp:2013vaa} is shown in Fig.~\ref{fig:event}(b). It may be noted here that the events distribution are obtained after incorporating the detector resolution as well as detector reconstruction efficiencies. 
\subsection{\label{subsec:binsch}{Binning scheme}}                    %
After incorporating the ICAL detector resolutions for muons and hadrons in 
neutrino induced events, a variable binning scheme has been adopted to study the 
sensitivity of sterile neutrinos oscillation. In a simulation the bin 
content may be less than one, but in an experiment, the observed data in 
various bins should be greater than or equal to one. The bins of different 
widths are chosen in order to ensure that there is at least one event in each 
bin. The binning schemes for down-going neutrinos induced events are given in
 Table 2 and 3. Table 2 gives the binning scheme for the down-going events 
using only the muon energy and zenith angle. Table 3 on the other hand gives 
the binning scheme for the analysis with the down-going events where the hadron 
energy information is used in addition to  muon energy and zenith angle. Since 
the muon reconstruction efficiency is  nearly zero for near-horizontal bin, we 
set a lower limit of $\cos\theta_{z} > 0.1$. The production of atmospheric 
neutrino flux follows a steep power law in energy($\sim$ E$^{-2.7}$), resulting 
in a smaller number of events at higher muon and hadron energies. Therefore, 
finer bins at low energies and wider bins at higher energies are considered for 
both muons and hadrons, respectively. Moreover, it may be noticed that at low 
 energies ($E = 1 - 1.5$ GeV) and high zenith angle ($\cos\theta_z = 0.1 - 
0.2$), a larger bin width is considered due to the fact that the reconstruction 
efficiency of ICAL is poor in this region. 

Similar arguments were adopted in choosing the binning scheme when considering 
events due to neutrinos coming from all directions. The details on the bin
combinations for the cases when only muon energy and its zenith angle are used 
and when hadron energy information is used as well, are given in Tables 4 and 5 
respectively. It is to be noted that wider bin combinations are considered for 
upcoming neutrinos induced events due to the further depletion of events as a 
result of oscillation of neutrinos in these cases. 

\begin{table}[h]
\caption{The binning scheme adopted for the reconstructed parameters E$_\mu$ and 
cos $\theta_{z}$ for muons induced due to downgoing neutrinos, where $\chi^{2}$ 
is estimated considering only muon events.}
\begin{tabular}{|c|c|c|c|}
\hline
Parameter & Range & Bin width & Total bins\\
\hline
\multirow{4}{40pt}{E$\mu$(GeV)} & [1, 1.5]  & 0.5 & 1 \\
 & [1.5, 3.0] & 0.25 & 6 \\
 & [3, 11] & 0.5 & 16\\
 & [11, 16] & 1 & 5\\
 & [16, 20] & 2 & 2\\
 \hline
cos $\theta_{z}$& [0.1, 0.2]& 0.05  & 2 \\
& [0.2, 1.0] & 0.025 & 32\\
\hline
\end{tabular}
\end{table}
\begin{table}[h]
\caption{The binning scheme adopted for the reconstructed parameters E$_\mu$, 
cos $\theta_{z}$ and E$_{h}$ for muons and hadrons, respectively, where 
$\chi^{2}$ is estimated considering combined muon and hadron information of 
downgoing neutrino induced events.}
\begin{tabular}{|c|c|c|c|}
\hline
Parameter & Range & Bin width & Total bins\\
\hline 
\multirow{4}{40pt}{E$\mu$(GeV)}& [1, 1.5] & 0.5 & 1\\
&[1.5, 5.5]&0.25&16\\ 
&[5.5, 8]&0.5&7\\
&[8, 13]&1&5\\
&[13, 17]&2&2\\
&[17, 20]&3&1\\
\hline
\multirow{2}{40pt}{cos $\theta_{z}$} & [0.1, 0.25] & 0.15&1\\
&[0.25, 1.0] &0.05 & 15 \\   
\hline
\multirow{2}{40pt}{E$_{h}$} & [0, 3] & 3&1\\
&[3, 20]&17&1\\
\hline  
\end{tabular}
\end{table}
\begin{table}[h]
\caption{The binning scheme adopted for the reconstructed parameters E$_\mu$ 
and cos $\theta_{z}$ for muons produced due to both upcoming and downgoing 
neutrinos, for the $\chi^{2}$ estimations with muons only.}
\begin{tabular}{|c|c|c|c|}
\hline
Parameter & Range & Bin width & Total bins\\
\hline
\multirow{4}{40pt}{E$\mu$(GeV)} & [1, 1.5]  & 0.5 & 1 \\
 & [1.5, 3.0] & 0.25 & 6 \\
 & [3.0, 6.0] & 0.5 & 6 \\
 & [6, 11] & 1.0 & 5\\
 & [11, 13] & 2 & 1\\
 & [13, 16] & 3 & 1\\
 & [16, 20] & 4 & 1\\
 \hline
cos $\theta_{z}$& [-1.0, -0.3]& 0.025  & 28 \\
& [-0.3, 0.1] & 0.4 & 1\\
& [0.1, 0.2] & 0.1 & 1\\
& [0.2, 1.0] & 0.025 & 32\\
\hline
\end{tabular}
\end{table}
\begin{table}[h]
\caption{The binning scheme adopted for the reconstructed parameters E$_\mu$, 
cos $\theta_{z}$ and E$_{h}$ for muons and hadrons where $\chi^{2}$ is estimated 
considering combined muon and hadron information and for the upcoming and 
downgoing neutrino induced events.}
\begin{tabular}{|c|c|c|c|}
\hline
Parameter & Range & Bin width & Total bins\\
\hline 
\multirow{4}{40pt}{E$\mu$(GeV)}& [1, 2] & 1.0 & 1\\
&[2, 7]&0.5&10\\ 
&[7, 10]&1.0&3\\
&[10, 12]&2&1\\
&[12, 15]&3&1\\
&[15, 20]&5&1\\
\hline
\multirow{2}{40pt}{cos $\theta_{z}$} & [-1.0, -0.3] & 0.1&7\\
&[-0.3, 0.1] &0.4 & 1 \\ 
&[0.1, 0.3] &0.1 & 2 \\
&[0.3, 1.0] &0.05 & 14 \\   
\hline
\multirow{2}{40pt}{E$_{h}$} & [0, 2] & 2&1\\
&[2, 20]&18&1\\
\hline  
\end{tabular}
\end{table}
\section{\label{sec:chi2esti}{\large{$\chi^{2}$} ESTIMATION}} %
To incorporate the muon energy and angle information in the analysis,  
we define a  $\chi^{2}$ as \cite{pu} 
\begin{equation}
\begin{split}
\chi^{2}_{muons} = \ & \min_{\xi_{i}}\sum_{n_1=0}^{N_1}\sum_{n_2=0}^{N_2} 
\left[ 2\left( R_{n_1,n_2}^{th}-R_{n_1,n_2}^{ex}\right)\right.\\
 & \left.+ 2 R_{n_1,n_2}^{ex} 
\,\ln\left(\frac{R_{n_1,n_2}^{ex}}{R_{n_1,n_2}^{th}}\right)\right]+\sum_{i=0}^{k
} \xi_{i}^{2}%
\end{split}%
\label{eq:chi1}
\end{equation}%
 and 
\begin{equation}
R_{n_1,n_2}^{th}=R_{n_1,n_2}^{'th}\left(1+\sum_{i=0}^{k}\pi_{n_1n_2}^{i}\xi_{i}
\right)+\mathcal{O}(\xi^{2})
\label{eq:chi2}
\end{equation}
where $n_1$ and $n_2$ are number of bins for energy and cosine of the zenith 
angle for muons, $R_{n_1,n_2}^{ex}$, $R_{n_1,n_2}^{'th}$ are observed and 
theoretically predicted events, $\pi_{n}^{i}$ is the strength of the coupling 
between the pull variable $\xi_{i}$ and $R_{n_1,n_2}^{th}$ which carries the 
information about systematic uncertainties as given in Eq.~(\ref{eq:chi2}). 
Equation~(\ref{eq:chi1}) is minimized with respect to pull variables. Five 
systematic uncertainties, $viz.$ an overall flux normalization error of 20$\%$ , 
overall normalization of cross-section of 10$\%$, flux tilt factor of 5$\%$
which takes into account the deviation of the atmospheric fluxes from a power 
law, zenith angle dependence of the flux of 5$\%$ and finally an overall 5$\%$ 
systematic error are considered.  More discussion on implementation of 
systematic uncertainty can be found in Ref.~\cite{Ghosh:2012px}. The 
information on the hadron energy is incorporated along with the muon energy and 
angle information by defining the $\chi^2$ as 
\begin{equation}
\begin{split}
\chi^{2}_{muons + hadrons} = \ & 
\min_{\xi_{i}}\sum_{n_1=0}^{N_1}\sum_{n_2=0}^{N_2}\sum_{n_3=0}^{N_3}\left[ 2 
\left( R_{n_1,n_2,n_3}^{th}- \right.\right. \\ \left. 
R_{n_1,n_2,n_3}^{ex}\right)\\
 & \left.+ 2 R_{n_1,n_2,n_3}^{ex} 
\,ln\left(\frac{R_{n_1,n_2,n_3}^{ex}}{R_{n_1,n_2,n_3}^{th}}\right)\right]+\sum_{
i=0 } ^ { k}\xi_{i}^{2}
\end{split}
\label{eq:chi3}
\end{equation}
where the index $n_3$ runs over the hadron energy bins. The other variables are 
the same as defined before. It may be noted that the theoretically predicted and 
observed events correspond to with and without sterile neutrino
oscillated events, respectively.
\begin{figure*}[]
\centering
{\includegraphics[width=0.4\linewidth,height=0.35\linewidth]{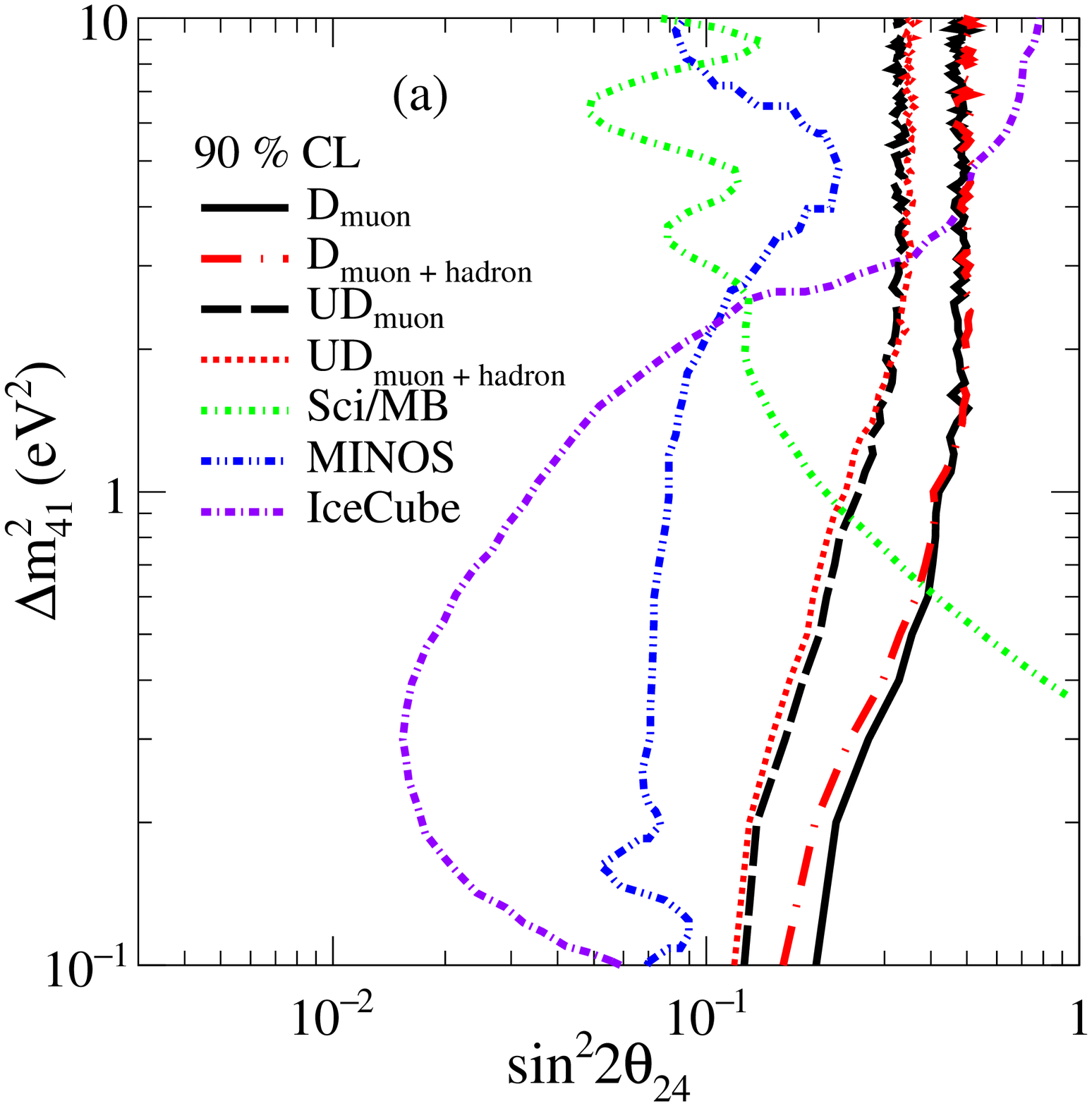}}
{\includegraphics[width=0.4\linewidth,height=0.35\linewidth]{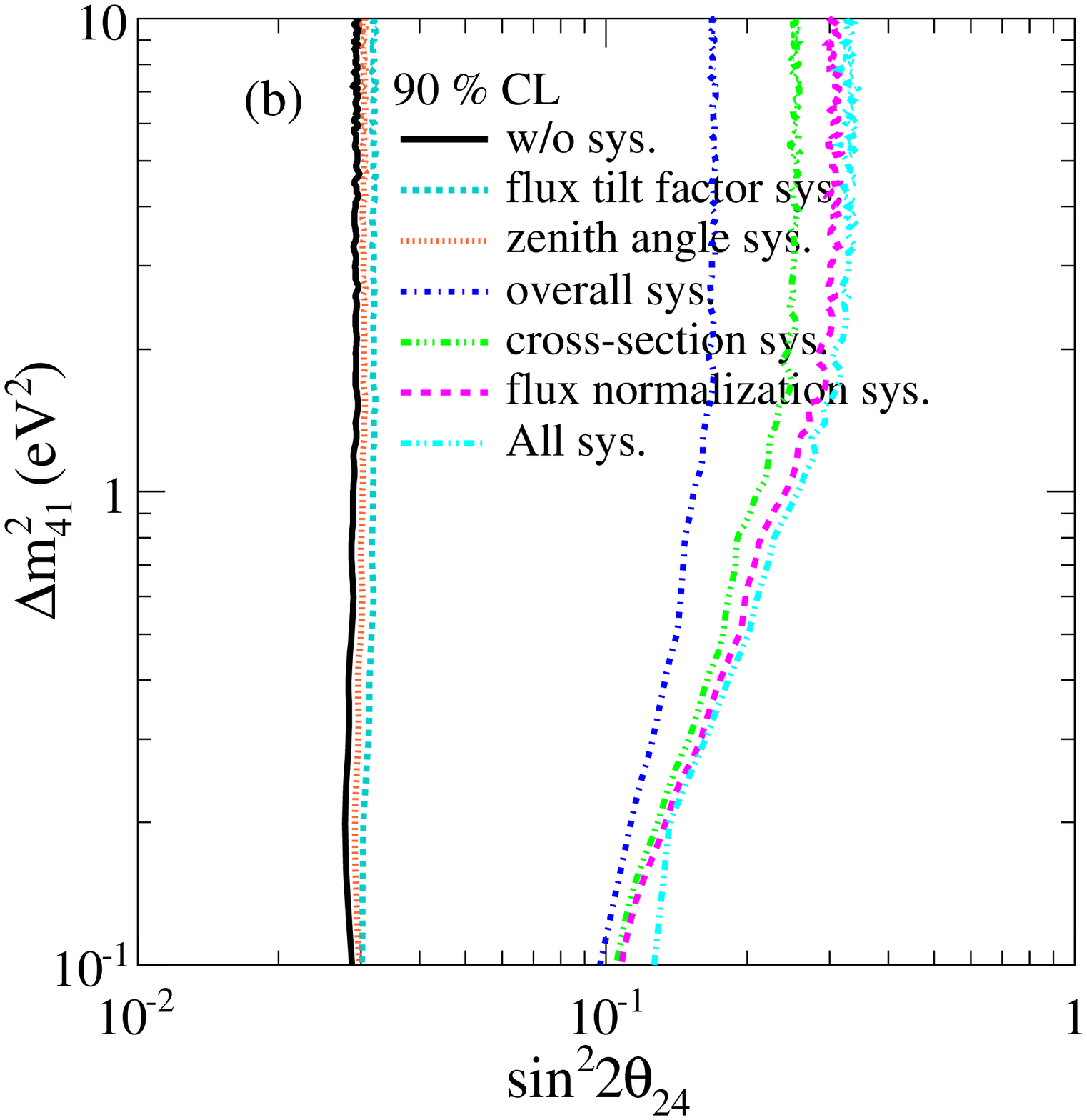}}
   \caption{\label{fig:exclusion}(a)The 90\% C.L. exclusion limits in the $\Delta 
m_{41}^2-\sin^22\theta_{24}$ plane, where $\sin^{2}2\theta_{24}= 4U^{2}_{\mu 
4}(1-U_{\mu 4})^{2}$, expected from 1 Mt-yr of the ICAL data. The legends `D' 
and `UD' correspond to only downward-going neutrinos and neutrinos from all 
directions. Here, the other active-sterile mixing angles are taken as 
$\theta_{14}$ = $\theta_{34}$ = 0.0$^{\circ}$. The 90\% C.L. SciBooNE/MiniBooNE, 
MINOS and IceCube exclusion regions are shown for comparison, (b) impact of individual systematic uncertainty on the sensitive limits.}
    \label{fig:fig1}
    \end{figure*}
The $\chi^2$ so defined is next marginalized over the sterile neutrino 
oscillation parameters  $\Delta m_{41}^{2}$ and the mixing angles $\theta_{14}$, 
 $\theta_{24}$ and $\theta_{34}$. The total $\chi^{2}$, is estimated by adding 
the priors of $\Delta m_{41}^{2}$ and $\theta_{14}$, $\theta_{24}$ and 
$\theta_{34}$,
\begin{equation}
\begin{split}
\chi^{2}_{total} =  \ & \chi^{2} + \left(\frac{(\Delta m^{2}_{41})^{bf} - \Delta m^{2}_{41})}{\sigma(\Delta m^{2}_{41})}\right)^{2} \\
& + \left(\frac{(\sin^{2}(2\theta_{ij})^{bf} - \sin^{2}(2\theta_{ij})}{\sigma(\sin^{2}(2\theta_{ij})}\right)^{2}
\end{split} 
\end{equation}
The parameters sin$^{2}2\theta_{ij}$ and $\Delta m^{2}_{41}$  are 
varied within the range of 0.047 to 0.22 and 0.82 eV$^{2}$ to 2.19 eV$^{2}$, 
respectively. The error on sin$^{2}2\theta_{ij}$ and $\Delta 
m^{2}_{41}$ are considered as 10$\%$ of their assumed true values. 
\section{\label{sec:excllimits}{EXCLUSION LIMITS}}    
\begin{figure*}[t!]
\centering
{\includegraphics[width=0.4\linewidth,height=0.35\linewidth]{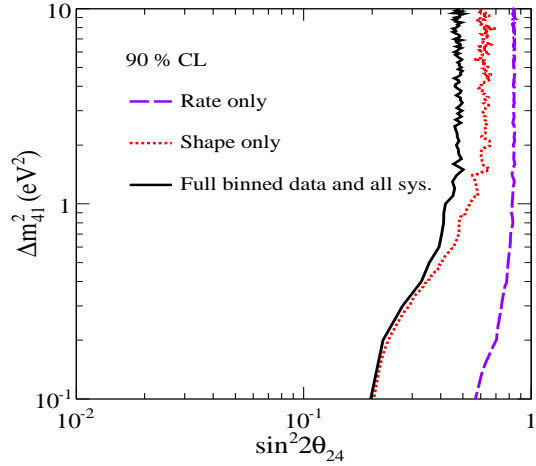}}
\caption{\label{fig:rateonly}Comparisons of exclusion limits in the 
$\Delta m_{41}^2$~-~$\sin^22\theta_{24}$ plane for rate-only, shape-only and including all systematics with consideration
of downgoing neutrinos only.}
    \label{fig:fig2}
    \end{figure*}
\begin{figure*}[t!]
\centering
{\includegraphics[width=0.4\linewidth,height=0.35\linewidth]{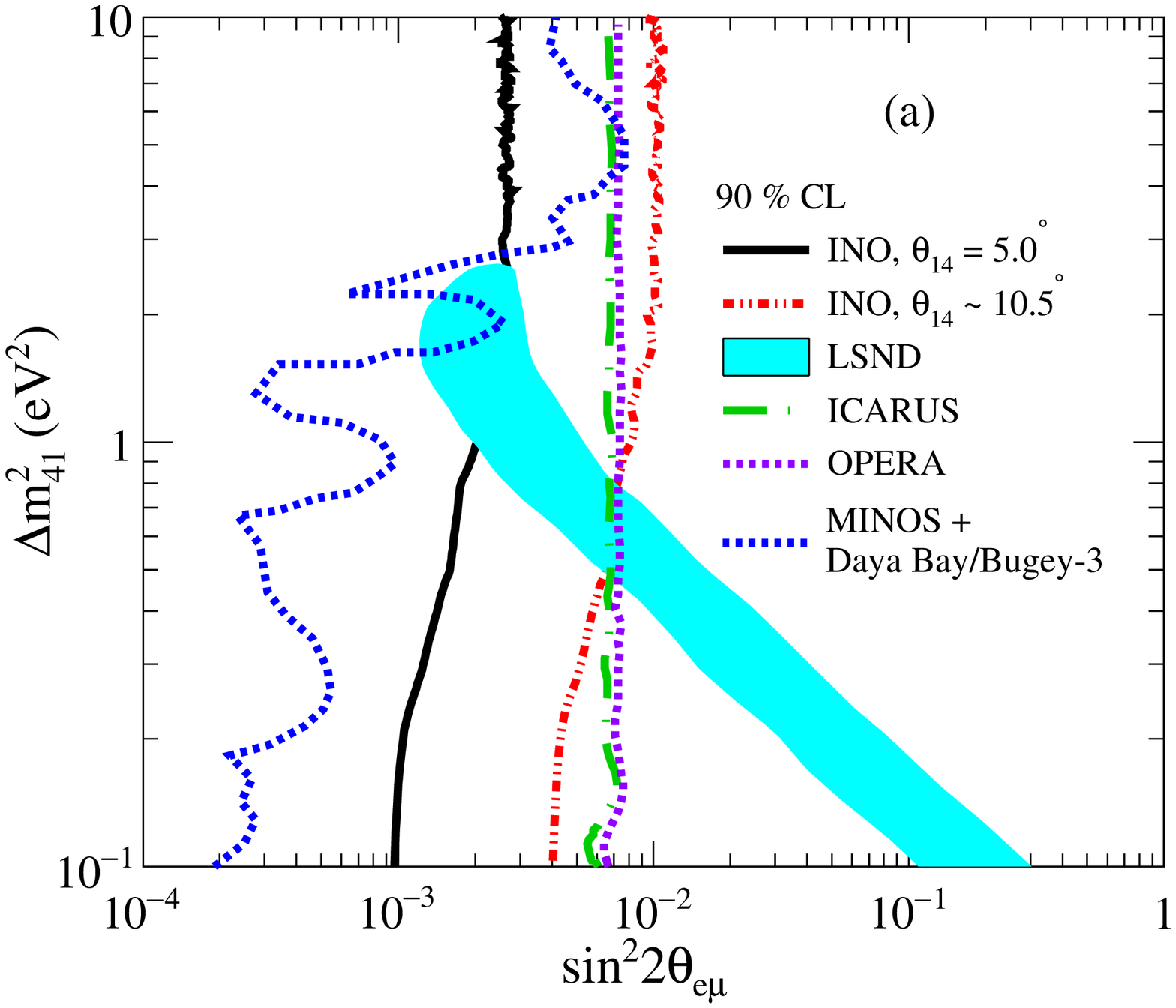}}
{\includegraphics[width=0.4\linewidth,height=0.35\linewidth]{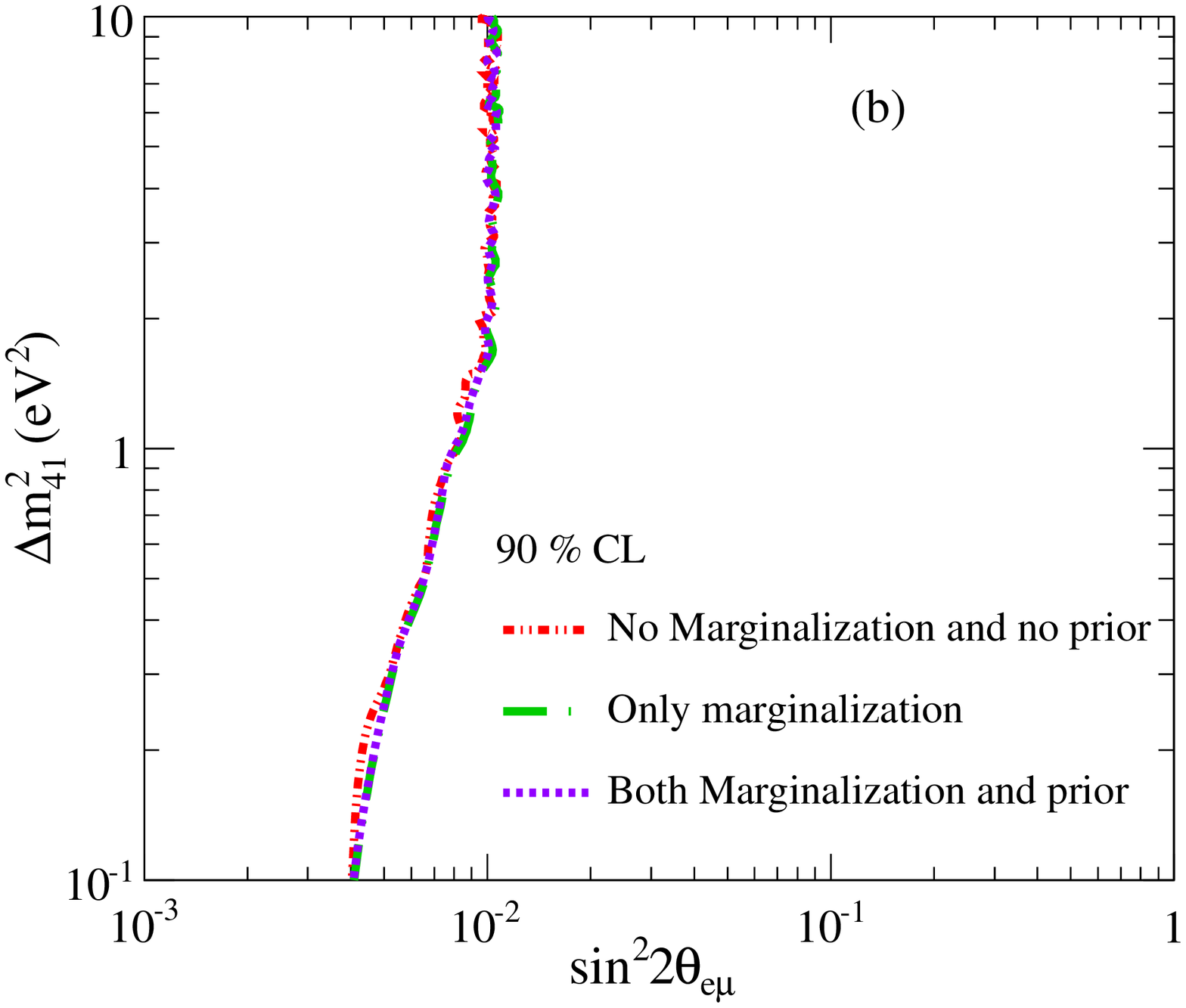}}
   \caption{\label{fig:excl2}(a)The 90\% exclusion limits in the 
$\Delta m_{41}^2$~-~$\sin^22\theta_{e\mu}$ plane where sin$^{2}2\theta_{e\mu}$= 
4 $U^{2}_{e 4}U_{\mu 4}^{2}$, expected from an exposure of 1 Mt-yr of the ICAL 
data. The active-sterile mixing angles are considered fixed at 
$\theta_{14}=5.0^{\circ}$ and 10.5$^{\circ}$ and $\theta_{34}$ = 0.0$^{\circ}$.
The 90\% C.L. allowed regions from the LSND and the 90\% C.L. excluded regions 
of OPERA, ICARUS and the combined results from MINOS and Daya Bay/Bugey-3 experiments are shown for comparisons, (b) effect of marginalization over $\theta_{14}$ with its 1$\sigma$ range.}
    \label{fig:fig2}
    \end{figure*}

The oscillation probabilities of atmospheric neutrinos depend on the 
active-sterile neutrino mixing. The data from ICAL will therefore be sensitive 
to this mixing and should be able to constrain them. The upper limit for the 
sterile neutrino mixing angle $\theta_{24}$ for an exposure of 1Mt-yr is shown 
in Fig.~\ref{fig:exclusion}. For simplicity we take the mixing angles 
$\theta_{14}$ and $\theta_{34}$ to be zero in this plot. The black lines 
correspond to the sensitivity obtained in the analysis when only muon energy and 
angle information is used, while the red lines correspond to the expected 
sensitivity when the hadron energy together with muon energy and angle 
information are included as well. A comparison of the black and red lines shows 
that the addition of hadron energy information in the analysis does not bring 
any significant improvement in the sensitivity. We show the expected sensitivity 
for the case where only down-going neutrinos are considered as well as when data 
from all zenith angles are analyzed. The lines labeled as `D' in the figure 
show the expected exclusion plots where only down-going events are considered,  
while the lines labeled as `UD' in the figure show the expected sensitivity 
when neutrino events from all zenith angles are included in the analysis. 
 The inclusion of up-going neutrinos is seen to improve the sensitivity to the 
 sterile mixing angle $\theta_{24}$. The improvement comes from the increase of 
 statistics as well as from the inclusion of earth matter effects which change with the 
 inclusion of active-sterile mixing. With 1 Mt-yr data, ICAL is expected to 
limit $\sin^{2}2\theta_{24}<0.19$ at the 90\% C.L. for $\Delta m_{41}^2 \sim 
0.1$ eV$^2$ using the down-going events only. This limit is expected to improve 
to $\sin^{2}2\theta_{24}<0.12$ when events from all zenith angles are included. 
This constitutes an improvement of about 35$\%$.

We also show in Fig.~\ref{fig:exclusion}(a), the exclusion plots from other 
experiments for comparison. The green dashed dotted line in 
Fig.~\ref{fig:exclusion}(a) shows the exclusion plot from the SciBooNE/MiniBooNE 
\cite{mb} experiment which looked for $\nu_{\mu}$ disappearance. It is 
found that at lower values of $\Delta m^{2}_{41}$, the ICAL detector has better 
sensitivity compared to SciBooNE/ MiniBooNE due to the longer path length and 
lower energies of the atmospheric neutrinos. The blue dashed double-dotted line 
shows results from the MINOS~\cite{Sousa:2015bxa} experiment for comparisons 
which corresponds to disappearance search of the $\nu_{\mu}$, in the range of 
$\Delta m^{2}_{43}$ = 0.1-10 eV$^{2}$ while their results spans over $\Delta 
m^{2}_{43}$ = 0.01-100 eV$^{2}$. In addition the violet line shows the result from IceCube~\cite{TheIceCube:2016oqi}. Figure~\ref{fig:exclusion}(b) shows the 
impact of individual systematics on the active-sterile mixing sensitivity.
It is found that the uncertainty due to flux normalization has the maximum effect
 compared to the others.
In addition the exclusion limits are obtained performing the rate only analysis considering single bin 
in energy and keeping the zenith angle bins same as given in Table~2 and also shape only analysis by 
removing the pull term on neutrino flux from the chi-square function where we consider the downgoing neutrinos only.
Figure~\ref{fig:rateonly} shows the comparison of exclusion limits
obtained from shape only (red dashed line) and the rate only (violet line) 
analysis at 90\% C.L. The lower limit on $\sin^{2}2\theta_{24} \sim$ 0.57 at $\Delta m^2_{41}$ = 0.1 eV$^2$ obtained from the
rate only analysis. However, while doing the rate only analysis by merging all the angle and energy bins, 
the limit on $\sin^{2}2\theta_{24} \sim$ 0.70 at $\Delta m^2_{41}$ = 0.1 eV$^2$
at 68~$\%$ C.L. It has been observed that the exclusion limits from shape only 
analysis ($\sin^{2}2\theta_{24} \sim$ 0.59) is less sensitive compare to the 
results obtained from the analysis of full binned data and including the pull corresponding to all systematic errors 
($\sin^{2}2\theta_{24} \sim$ 0.47) at higher values of $\Delta m^2_{41}$ = 10.0 eV$^2$. 
However, at low value of $\Delta m^2_{41}$ = 0.1 eV$^2$ the difference is $\sim$ 3~$\%$.    
\begin{figure*}[t!]
 \includegraphics[width=0.32\linewidth,height=0.32\linewidth]{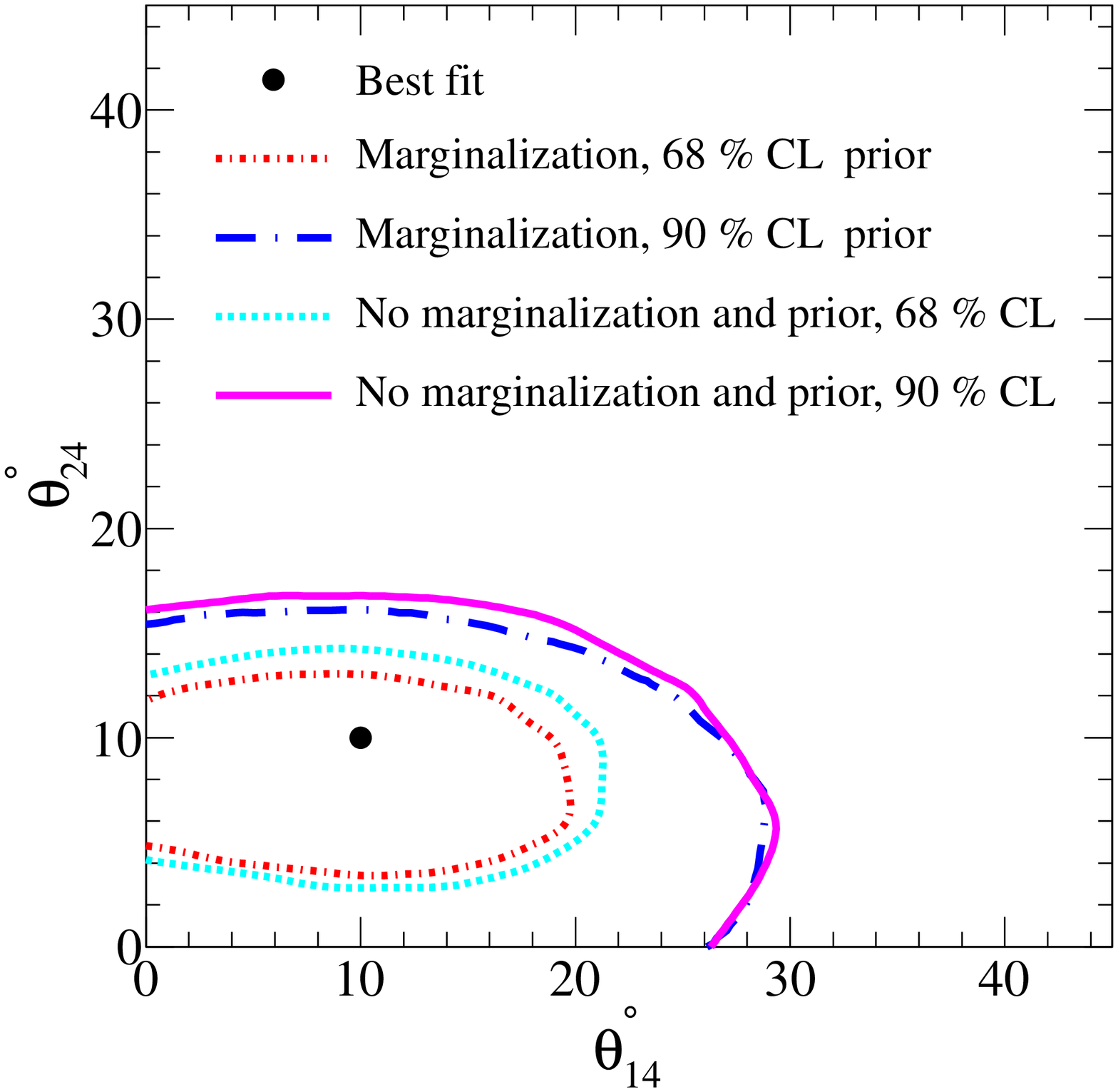}
  \includegraphics[width=0.32\linewidth,height=0.32\linewidth]{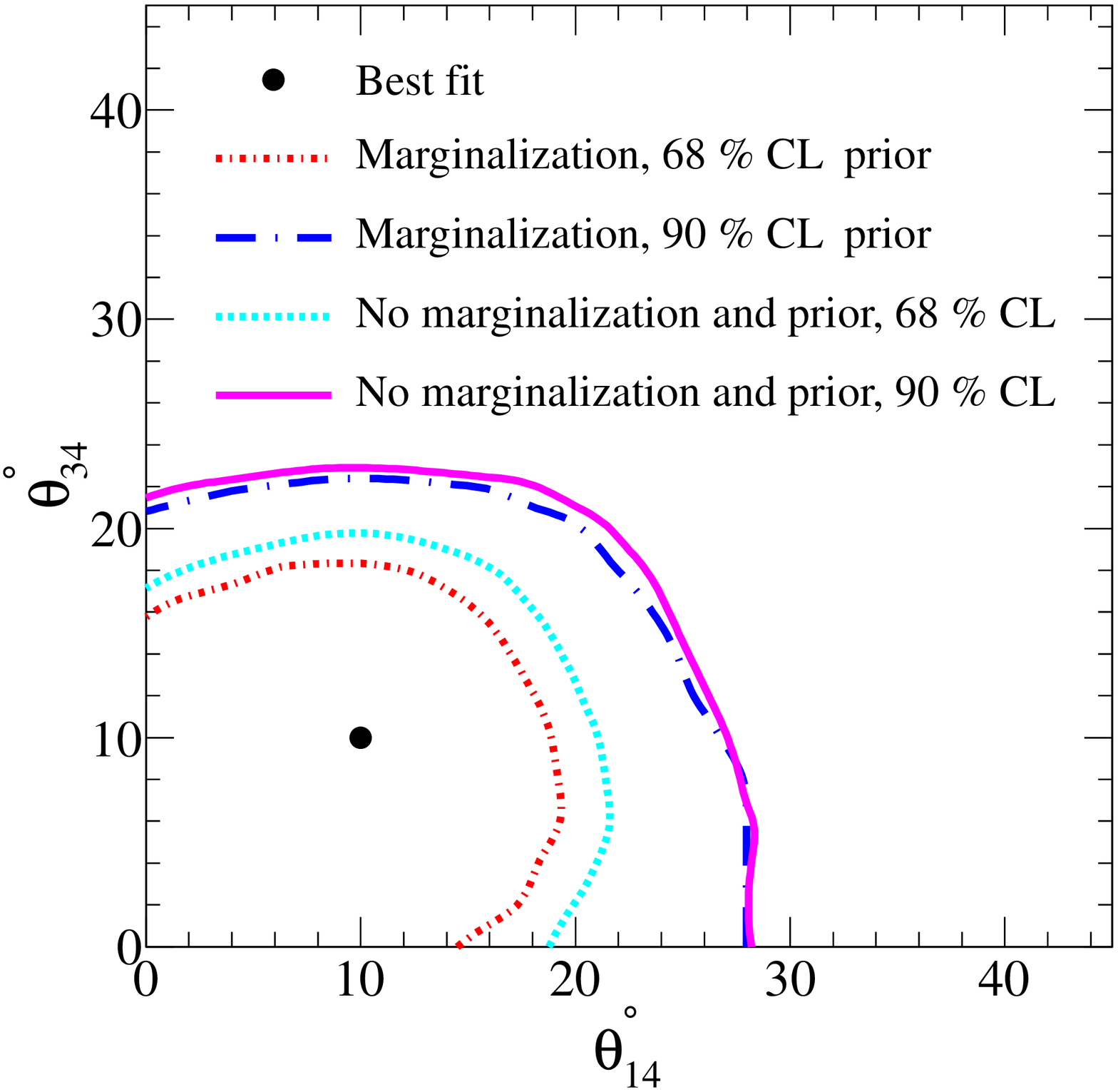}
  \includegraphics[width=0.32\linewidth,height=0.32\linewidth]{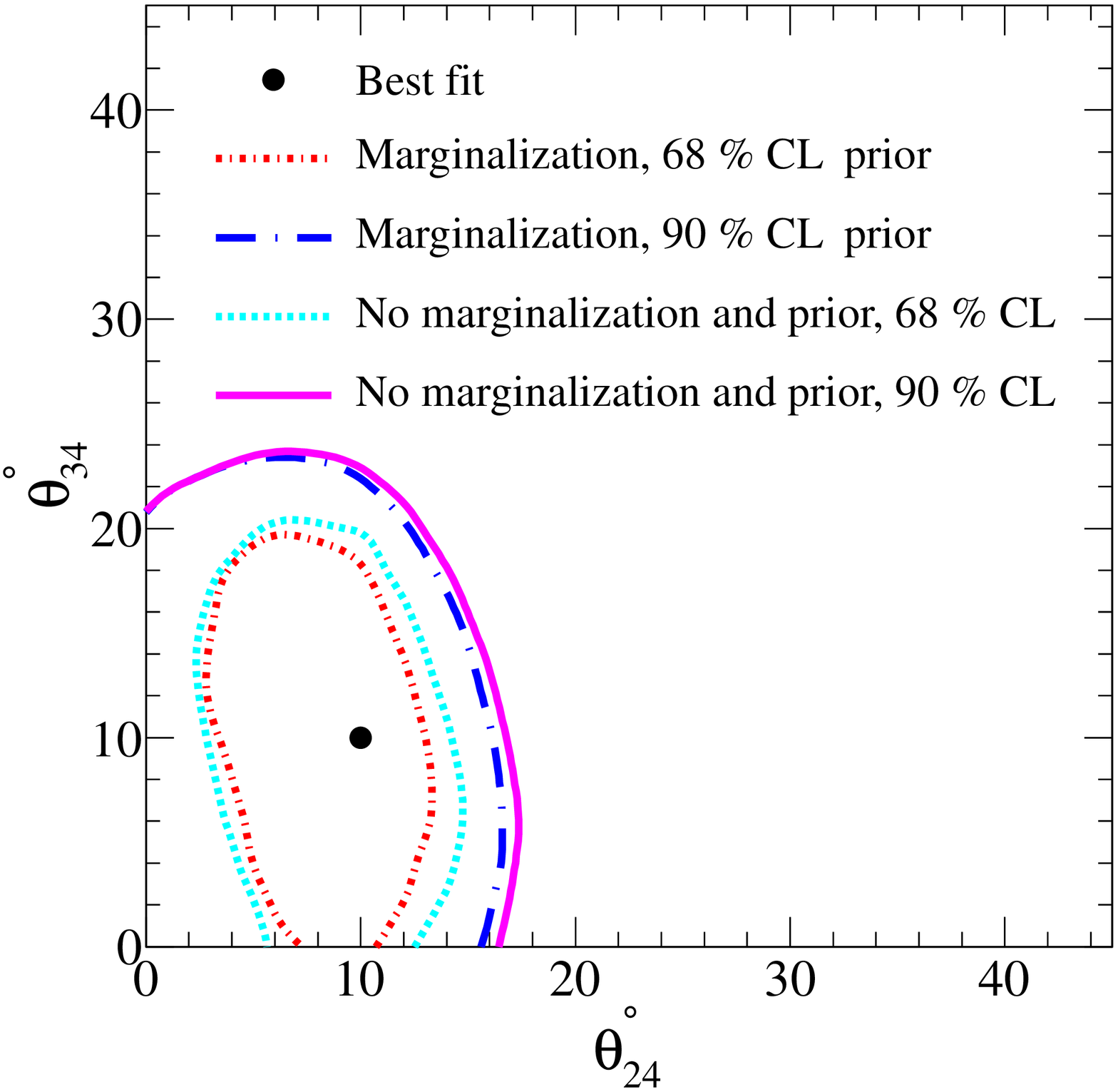}
\caption{ \label{fig:excl34} The expected allowed areas in the 
$\theta_{14}-\theta_{24}$ plane (left-hand panel), $\theta_{14}-\theta_{34}$ 
plane (middle panel), and $\theta_{24}-\theta_{34}$ plane (right-hand panel), 
expected from 1 Mt-yr of ICAL data. The standard 3-generation oscillation 
parameters are taken as sin$^{2} \theta_{12}=0.3$, sin$^{2} 2\theta_{13}= 0.1$, 
sin$^{2}\theta_{23}=0.5$ , $\Delta m^{2}_{21}$ = 7.5 $\times$ 10$^{-5}$ eV$^{2}$ 
, $\Delta m^{2}_{31}$ = 2.4 $\times$ 10$^{-3}$ eV$^{2}$. The data is generated 
at $\theta_{14} = \theta_{24} = \theta_{34} = 10^{\circ}$ and at $\Delta 
m^{2}_{41} = 1$ eV$^{2}$. We marginalize the $\chi^2$ over $\Delta m^{2}_{41}$ 
and the remaining sterile mixing angle, {\it viz.}, $\theta_{34}$ in the 
left-hand panel, $\theta_{24}$ in the middle panel and $\theta_{14}$ in the 
right-hand panel. The priors considered are described in Sec.~\ref{sec:chi2esti}.}
    \end{figure*}
Figure~\ref{fig:excl2}(a) shows the expected sensitivity of ICAL to the 
effective mixing angle $\theta_{e\mu}$ which for the active-sterile mixing case 
is defined as $\sin^{2}2\theta_{e\mu}= 4 U^{2}_{e 4}U_{\mu 4}^{2}$. We have 
taken a statistics corresponding to 1 Mt-yr data  at ICAL and considered the 
information on just the muon energy and zenith angles. We show the 90\% C.L. 
expected exclusion limits in the $\Delta m^{2}_{41} - \sin^{2}2\theta_{e\mu}$ 
plane  for fixed choices of $\theta_{14}\sim 5.0^{\circ}$ (solid black 
line) and $10.5^{\circ}$ (dashed double-dotted red line). For both these case we 
keep $\theta_{34} = 0.0^{\circ}$. The cyan shaded region shows the 90\% C.L. 
allowed region from the LSND~\cite{Aguilar:2001ty} experiment for comparison. 
The green long-dashed line shows the results from the 
ICARUS~\cite{Antonello:2013gut} while the purple dotted line shows the results 
OPERA~\cite{Agafonova:2013xsk}. The sensitivity of the ICAL experiment is seen 
to increase significantly depending on the value of $\theta_{14}$ and could in 
principle rule out significant parts of the LSND allowed region.
The blue dashed line shows results obtained from combined analysis of MINOS and Daya Bay/Bugey-3 experimental data at 90\% C.L~\cite{Adamson:2016jku}. Figure~\ref{fig:excl2}(b)
shows the marginalization effect of $\theta_{14}$ on the sensitivity $\sin^{2}2\theta_{e\mu}$.
The 1$\sigma$ range of $\theta_{14}$ was considered from Ref.~\cite{Giunti:2013waa}.
It is observed that the impact on $\sin^{2}2\theta_{e\mu}$ is negligible.  
Further exclusion plots are generated for various mixing angle combinations 
considering only muon energies and zenith angles and shown in 
Fig.~\ref{fig:excl34}. The data for these figures are generated for assumed 
true values of sterile mixing angles of $\theta_{14} = \theta_{24} = \theta_{34} 
= 10^{\circ}$ and at $\Delta m^{2}_{41}$ = 1 eV$^{2}$. The assumed true values 
of the standard oscillation parameters are given in the figure caption. 
Figure~\ref{fig:excl34} shows the expected constraints from 1 Mt$-$yr of ICAL 
data in the $\theta_{14}-\theta_{24}$ plane in the left-hand panel, 
$\theta_{14}-\theta_{34}$ plane in the middle-panel, and 
$\theta_{24}-\theta_{34}$ plane in the right-hand panel. The point at which the 
data is generated is shown by the black dots. The different lines show the 
different marginalizing methods, with added priors and at different 
C.L., and the details are given in the figure legends. For the cases which 
include marginalization, we have marginalized the $\chi^2$ over $\Delta 
m^{2}_{41}$ and the remaining sterile mixing angle, {\it viz.}, $\theta_{34}$ in 
the left-hand panel, $\theta_{24}$ in the middle panel and $\theta_{14}$ in the 
right-hand panel, with priors, as was discussed in the previous 
section. For the case with marginalization, we expect an upper bound at 90\% 
C.L. (from 2 parameter plots) of around $20^\circ$ for $\theta_{14}$ and 
$\theta_{34}$, and about  $12^\circ$ for $\theta_{24}$. On the lower side only 
$\theta_{24}$ is seen to be bounded at 90\% C.L. without priors and at 68\% 
C.L. once priors are included. While for both the other mixing angles, the 
cases $\theta_{14}=0$ and $\theta_{34}=0$ are expected to be compatible with the 
data at even the $1\sigma$ C.L. for all analyses that we considered.

\section{\label{sec:summ}{SUMMARY}}              %
The sensitivity of ICAL to active-sterile neutrino mixing is studied. The effect 
of inclusion of hadron energy into the analysis has been probed. The present 
analysis shows that the inclusion of upcoming neutrinos, which are affected by 
the intervening matter, improves the sensitivity to sterile neutrino mixing. The 
inclusion of hadron energy information in the analysis, on the other hand, has 
almost no effect on the sensitivity to sterile neutrino mixing. From the 
down-going events alone, one expects an upper bound of $\sin^22\theta_{24} < 
0.16$ at 90\% C.L. from 1 Mt-yr of data. The sensitivity to the sterile 
neutrino mixing angle further improves by considering neutrinos coming from all 
directions at the detector. There is enhancement in sensitivity by about 35$\%$ 
for the whole range of $\Delta m^2_{41}$. At lower values of $\Delta m^2_{41}$, 
the ICAL detector has better sensitivity compared to the short baseline 
experiments like SciBooNE/MiniBooNE. Also expected bounds on the sterile mixing 
angles are obtained. For an illustrative case of assumed true value of 
$10^\circ$ for all the three sterile mixing angles and after marginalization, 
an upper bound at 90\% C.L. (from 2 parameter plots) of around $20^\circ$ for 
$\theta_{14}$ and $\theta_{34}$, and about  $12^\circ$ for $\theta_{24}$ is 
obtained. The impact of inclusion of priors on the sterile neutrino parameters 
is studied. Only for $\theta_{24}$ could one rule out the zero-mixing 
possibility  at 90\% C.L. for this illustrative case, while for both the other 
mixing angles $\theta_{14}=0$ and $\theta_{34}=0$ are compatible with the data 
at even the $1\sigma$ C.L..   
\\
 
\section*{ACKNOWLEDGMENTS}
We thank the INO Collaboration for useful suggestions and discussions.
We also thank Vibhuti Duggal for her support while running the program at the BARC supper-computing facility.
We would also like to thank S. Umasankar, A. Raychaudhuri, S. Goswami and  Md. Naimuddin for their helpful suggestions 
and useful discussions. SC acknowledges support from the Neutrino Project under the XII plan of Harish-Chandra Research Institute and partial support from the European Union FP7 ITN INVISIBLES (Marie Curie Actions, PITN-GA-2011-289442).


\end{document}